%% file: SUS-10-008_temp.tex
\pdfoutput=1

\documentclass[11pt,twoside,a4paper,cmspaper,final,collab]{cms-tdr}
\def\svnVersion{80618}\def\cmsCernNo{2011-046}\def\cmsCernDate{\today}\def\cmsMessage{Submitted to Physics Letters B}

\begin{document}\cmsNoteHeader{SUS-10-008}

\hyphenation{had-ron-i-za-tion}
\hyphenation{cal-or-i-me-ter}
\hyphenation{de-vices}
\RCS$Revision: 79081 $
\RCS$HeadURL: svn+ssh://alverson@svn.cern.ch/reps/tdr2/papers/SUS-10-008/trunk/SUS-10-008.tex $ \RCS$Id: SUS-10-008.tex 79081 2011-09-17 22:53:57Z chertok $
\newcommand{\jt}{\ensuremath{H_{\mathrm{T}}}\xspace}
\newcommand{\GeVC}{\ensuremath{\GeV\!/c}}
\newcommand {\abseta} {|\eta|}
\newcommand{\mzero}{\ensuremath{{m_0}}}
\newcommand{\azero}{\ensuremath{A_{0}}}
\newcommand{\mhalf}{\ensuremath{m_{1/2}}}
\newcommand{\tb}{\ensuremath{\tan\beta}}
\newcommand{\nall}{\tilde{{\chi}}^{0}}
\newcommand{\call}{\tilde{{\chi}}^\pm}
\newcommand{\ntwo}{\tilde{{\chi}}^{0}_{2}}
\newcommand{\cone}{\tilde{{\chi}}^\pm_{1}}
\newcommand{\none}{\tilde{{\chi}}^{0}_{1}}
\newcommand{\nunit}[2]{#1\,\mbox{#2}}
\newcommand{\met}{\ensuremath{E_{\mathrm{T}}^{\text{miss}}}\xspace}
\newcommand{\curlumi}{35}
\ifthenelse{\boolean{cms@external}}{\newcommand{\xright}{bottom\xspace}}{\newcommand{\xright}{right\xspace}}
\ifthenelse{\boolean{cms@external}}{\newcommand{\xleft}{top\xspace}}{\newcommand{\xleft}{left\xspace}}
\ifthenelse{\boolean{cms@external}}{\newcommand{\xRight}{Bottom\xspace}}{\newcommand{\xRight}{Right\xspace}}
\ifthenelse{\boolean{cms@external}}{\newcommand{\xLeft}{Top\xspace}}{\newcommand{\xLeft}{Left\xspace}}
\cmsNoteHeader{SUS-10-008}

\title{Search for Physics Beyond the Standard Model Using Multilepton Signatures in
\Pp\Pp\ Collisions at $\sqrt{s}=7\TeV$ }
\date{\today}

\abstract{

A search for physics beyond the
standard model in events with at least three leptons and any
number of jets is presented.
The data sample corresponds to 35~pb$^{-1}$ of
integrated luminosity in \Pp\Pp\ collisions at
$\sqrt{s}=7$~TeV
collected by the CMS experiment at the LHC.
A number of exclusive multileptonic channels are
investigated and standard model backgrounds are
suppressed by requiring sufficient missing transverse energy, invariant mass
inconsistent with that of the \textrm{Z} boson, or high jet activity.
Control samples in data are used to ascertain the robustness
of background evaluation techniques and to minimise the reliance
on simulation.  The observations are consistent with background
expectations. These results constrain previously
unexplored regions of supersymmetric parameter space.

}
\hypersetup{%
pdfauthor={CMS Collaboration},%
pdftitle={Search for Physics Beyond the Standard Model Using Multilepton Signatures in pp Collisions at sqrt(s)=7 TeV},%
pdfsubject={CMS},%
pdfkeywords={CMS, physics, supersymmetry, multileptons, tau, mSUGRA, RPV, GMSB}}

\maketitle 
\section{Introduction}
\label{sec:intro}

Supersymmetry (SUSY) is a preferred candidate for a theory beyond the
standard model (SM) because it solves the hierarchy problem, allows the
unification of the gauge couplings, and may provide a candidate particle for
dark matter~\cite{Nilles:1983ge,Haber:1984rc,deBoer:1994dg}.
The 7~TeV centre-of-mass energy of the Large Hadron Collider (LHC)
makes it possible to search for squark and gluino production in
previously unexplored regions of supersymmetric parameter space with
the integrated luminosity delivered in the first few months of operation.
Hadronic collisions yielding three or more electrons, muons, or taus
(``multileptons'') serve as an ideal hunting ground for physics beyond
the SM, as leptonic SM processes are relatively rare at
hadron colliders and multilepton events particularly so.

We report results from a search with broad sensitivity to the
potentially large multilepton signals from SUSY
particle production. Our strategy takes advantage of the strong
background suppression obtained when requiring three or more leptons;
this allows us to relax requirements for SM background reduction
relative to other searches with fewer leptons or purely hadronic searches at
the LHC~\cite{Khachatryan:2011tk,PhysRevLett.106.131802}.

The multilepton search presented here is not tailored for any
particular SUSY scenario.  Nonetheless, it probes multiple new regions
of the supersymmetric parameter space beyond previous multilepton
searches at the
Tevatron~\cite{Aaltonen:2008pv,Dube:2008kf,Ruderman:2010kj,Abazov:2009zi,Forrest:2009gm,Abazov:2006nw,Abulencia:2007mp}.
Overall, this search
complements the Tevatron searches, which are mostly sensitive
to electroweak gaugino production,
while this search is mostly sensitive to squark-gluino production.
As in the case of Tevatron searches, we interpret results in the
mSUGRA/CMSSM~\cite{MSUGRA,CMSSM} scenario of supersymmetry in which
the superpartner masses and gauge couplings become unified at the grand
unification scale, resulting in common masses $m_0$ ($m_{1/2}$) for
all spin 0 (1/2) superpartners at this scale.  The remaining CMSSM
parameters are $\azero,\tb$, and $\mu$. For illustration,
we define a CMSSM benchmark point called ``TeV3'', characterised by
$\mzero=60\GeVcc,\mhalf=230\GeVcc,\azero=0,\tb=3,\mu>0$, and a
next-to-leading order (NLO)
cross section of 10~pb for all supersymmetric processes.

In this article we also study scenarios with gravitinos as the
lightest supersymmetric particle (LSP) and
sleptons as the next-to-lightest supersymmetric particles (NLSPs).
Scenarios of this type arise in a wide class of theories of gauge
mediation with split messengers
(GMSM)~\cite{Dimopoulos:1996va,Culbertson:2000am}.  Multilepton final
states arise naturally in the subset of the GMSM parameter space
where the right-handed sleptons are flavour-degenerate, the so-called
slepton co-NLSP
scenario~\cite{Dimopoulos:1996va,Culbertson:2000am,Ruderman:2010kj,
Alves:2011wf}.
We define a slepton co-NLSP benchmark point, called ML01, characterised by
a chargino mass $m_{\chi^{\pm}}=385\GeVcc$ and gluino mass
$m_{\tilde{g}}=450\GeVcc$.
The other superpartner masses are then
given by the generic relationships
$m_{\tilde{\ell}_R}=0.3m_{\chi^{\pm}},
m_{\chi^0_1}=0.5m_{\chi^{\pm}},
m_{\tilde{\ell}_L}=0.8m_{\chi^{\pm}}$, and
$m_{\tilde{q}_L}=0.8m_{\tilde{g}}$.
ML01 has an estimated 45~pb NLO cross section for all supersymmetric
processes.  Finally, we also consider the possibility that the LSP is
unstable.

\section{Detector}

The data sample used in this search corresponds to the integrated
luminosity of \curlumi\ \pbinv\ recorded in 2010 with the Compact Muon
Solenoid (CMS) detector at the LHC, running at 7~TeV centre-of-mass
energy.  The CMS detector has cylindrical symmetry around
the \Pp\Pp\ beam axis with tracking and muon detector
pseudorapidity coverage to $\abseta<2.4$, where $\eta = - \ln \tan
(\theta/2)$ and $\theta$ is the polar angle with respect to the
counterclockwise beam.  The azimuthal angle $\phi$ is measured in
the plane perpendicular to the beam direction.  Charged particle
tracks are identified with a 200 m$^2$, fully silicon-based tracking
system composed of a pixel detector with three barrel layers at radii
between 4.4 cm and 10.2 cm and a silicon strip tracker with 10 barrel
detection layers, of which four are double sided, extending outwards
to a radius of 1.1 m. Each system is completed by endcaps extending
the acceptance of the tracker up to a pseudorapidity of
$\abseta<2.5$. The lead-tungstate scintillating crystal
electromagnetic calorimeter (ECAL) and brass/scintillator hadron
calorimeter hermetically surrounding the tracking system measure the
energy of showering particles with $\abseta<3.0$. These
subdetectors are placed inside a 13 m long and 6 m diameter
superconducting solenoid with a central field of 3.8 T. Outside the
magnet is the tail-catcher of the hadronic calorimeter followed by the
instrumented iron return yoke, which serves as a multilayered muon
detection system in the range $\abseta< 2.4$.  The CMS detector has
extensive forward calorimetry, extending the pseudorapidity coverage
to $\abseta<5.0$.  The performance of all detector components
as measured with cosmic rays has been reported in Ref.~\cite{Chatrchyan:2009hb} and references therein.
A much more detailed description of CMS can be found elsewhere~\cite{:2008zzk}.

\section{Event Trigger}

The data used for this search came from  single- and double-lepton triggers. The Level-1 (L1) and  High Level Trigger (HLT)
configurations of the CMS trigger were adapted
to changing beam conditions and increasing LHC luminosities during data collection.
For example, the transverse momentum (\pt\ ) threshold for the
unprescaled single muon trigger was raised
from 9\GeVc to 15\GeVc near the end of data
taking. The analogous single electron trigger went from
a transverse energy (\et) threshold of 10 GeV in the early part of data taking to 17
GeV.  Double-lepton trigger thresholds were set at  $\pt > 5\GeVc$
for muons and $E_{\mathrm{T}} > 10 \GeV$ for electrons.

The efficiencies of the single-lepton triggers are determined
with the tag-and-probe technique.  Events with \textrm{Z} boson decays into
two electrons or muons are selected by requiring one lepton and another
track as a lepton candidate, with an invariant mass in the \textrm{Z}-mass
window of 80 to $100\GeVcc$. The fraction of probed tracks that are
reconstructed correctly as leptons including the trigger requirements
determines the lepton efficiency. The average trigger efficiency
determined for $\pt > 15\GeVc$ is
$97.5 \pm 1.5\%$ for the electrons and $89.1 \pm 0.9\%$ for the muons.
\section{Lepton Identification}
Leptons in this search can be either electrons, muons, or taus.
Electrons and muons are selected
with $\pt \geq 8\GeVc$ and $|\eta| < 2.1$ as reconstructed from
measured quantities from the tracker, calorimeter, and muon system.
Since a large fraction of the data set was collected with the highest
trigger threshold implemented at high luminosity, we require at least
one identified muon with $\pt > 15\GeVc$ or an electron with
$E_{\mathrm{T}}>20$~GeV.
The matching candidate tracks must satisfy quality requirements and
spatially match with the energy deposits in the ECAL and the tracks in
the muon detectors, as appropriate.  Details of reconstruction and
identification can be found in Ref.~\cite{EGM-10-004} for electrons
and in Ref.~\cite{MUO-10-002} for muons.  Jets are reconstructed using
particles with $\abseta \leq 2.5$ via the particle-flow (PF) algorithm, as
described in Ref.~\cite{PFT-10-002}.

Although the reconstruction of taus presents challenges, we include
these because there are regions of parameter space where
signatures that include taus are enhanced.  Taus decay either
leptonically or hadronically. The electrons or muons from the leptonic
decays are identified as above. The hadronic decays yield either a
single charged track (one-prong decays) or three charged tracks
(three-prong decays) with or without additional electromagnetic energy
from neutral pion decays. We explore two strategies for hadronic decay
reconstruction in this search and combine the results in the end.  In
the first selection, the one-prong hadronic decays are reconstructed
as isolated tracks with $p_{\mathrm{T}}>8\GeVc$.
In the second selection, hadronic decays are reconstructed with the
PF algorithm~\cite{PFT-10-004,PFT-08-001}, which also
includes the three-prong decays and decays with associated ECAL
activity.  This algorithm defines an energy-dependent signal cone in
the $\eta$-$\phi$ region around the candidate track with an angular radius
$\Delta R=\sqrt{(\Delta\eta)^2+(\Delta\phi)^2}$
of $5\GeV /\et(\mathrm{jet})$.  This ''shrinking cone'' is limited to the
range $0.07\leq \Delta R \leq 0.15$.
Inside the signal cone one or three charged tracks are required. PF
tau candidates that are also
electron or muon candidates are explicitly rejected.

These two algorithms have complementary benefits. Isolated tracks originating from one-prong decays make up only about 18\% of hadronic tau decays, but have relatively low backgrounds.  Additionally, some electrons and muons that fail normal requirements described above are accepted with the isolated track reconstruction.
The PF algorithm reconstructs all hadronic tau decays including the larger-background
three-prong decays, necessitating tighter kinematic requirements for some event
topologies.  After event selection, the tau channel efficiencies are similar for both selections.

Sources of
background leptons include genuine leptons occurring inside or near
jets, hadrons simulating leptons by punch-through into the muon
system, hadronic showers with large electromagnetic fractions, or
photon conversions.  An isolation requirement strongly reduces the
background from misidentified leptons, since most of them occur inside
jets.  We define the relative isolation $I_{\mathrm{rel}}$ as the ratio of the
sum of calorimeter energy and $\pt$ of any other tracks in the cone
defined by $\Delta R < 0.3$ around the lepton to the $\pt$ of the
lepton.  For electrons, muons, and isolated tracks, we require
$I_{\mathrm{rel}}<0.15$. For PF taus, tracking and ECAL
isolation requirements are applied~\cite{PFT-08-001} in the annular
region between the signal cone and an isolation cone with $\Delta R =
0.5$.

Leptons from SUSY decays considered in this search
originate from
the collision point ("prompt" leptons).
After the isolation selection, the most significant background sources
are residual nonprompt leptons from heavy quark decays, where
the lepton tends to be more isolated because of the high \pt\ with
respect to the jet axis.  This background is reduced by requiring that
the leptons originate from within one centimeter of the primary vertex
in $z$ and that the
impact parameter $d_{xy}$ between the track and the event vertex in
the plane transverse to the beam axis be small. For electrons,
muons, and isolated tracks, the impact parameter requirement is
$d_{xy}\le 0.02$ cm, while $d_{xy}\le 0.03$ cm is required for
PF taus.
The isolation and promptness criteria are efficient for the SUSY
signal but almost eliminate misidentified leptons.

\section{Search Strategy}

\subsection{Multilepton channels}

Candidate events in this search must have at least three leptons, of
which at least one must be an electron or a muon, and may contain two or fewer hadronic tau candidates.  We classify
multilepton events into search channels on the basis of the number of
leptons, lepton flavour, and relative charges as well as charge and
flavour combinations and other kinematic quantities described below.

We use the following symbols and conventions in describing the
search. The symbol $\ell$ stands for an electron or a muon, including
those from tau decays.  In describing pairs of leptons,
OS stands for opposite-sign, SS for same-sign, and SF for
same (lepton) flavour. To explicitly denote differing lepton flavours in a pair,
we use the symbol $\ell\ell'$.  
The symbol $\tau$ refers to hadronic tau decays reconstructed using the PF tau
algorithm and $T$ refers to decays reconstructed as isolated tracks.

The level of SM background varies considerably across the channels.
Channels with hadronic tau decays or containing OS--SF
($\ell^{\pm}\ell^{\mp}$) pairs
suffer from large backgrounds, but channels such as
$\ell^{\pm}\ell^{\pm}\ell'$ have smaller backgrounds because they do not contain
OS--SF pairs.
High-background channels play two distinct roles in this search, depending on the scenario of new physics.  For models that predict a small signal yield in these channels, they act as "control" samples that give confidence in predictions for the ÒdiscoveryÓ channels that have small background.  But it is also possible that new physics may preferentially manifest itself 
in the high-background channels. For example, taus can greatly outnumber electrons and muons in the case of supersymmetry 
with large $\tan\beta$ values. Therefore, we retain channels such as those with two hadronic tau decays although they contribute only modestly to scenarios of new physics which we discuss later. In comparison, dilepton searches have higher backgrounds and are thus less sensitive to tau-rich signals because of additional requirements necessary to reduce these backgrounds to a manageable level.  We avoid using kinematic quantities such as \met\ or \jt\ in defining datasets used in background determination.  Such loose selection criteria minimize signal contamination between high and low background channels.  Even for tau-rich mSUGRA scenarios, the signal contamination is below 5\%.

\subsection{Background reduction}

Other searches for new physics such as those requiring dileptons or
single leptons suffer from large SM backgrounds and are hence
forced to require substantial jet activity
as well as missing transverse energy. For the multilepton search
described here, the presence of a third lepton
results in lower SM backgrounds, thus reducing
reliance on other requirements and increasing sensitivity to diverse
signatures of new physics.
The presence of hadronic activity in an
event is characterised by the variable \jt, defined as the scalar sum
of the transverse jet energies for all jets with $\et>30~\GeV$ and $\abseta<2.4$.  Jets
used for the \jt\ determination must be well separated from any
identified leptons; jets are required to have no lepton in a cone
$\Delta R < 0.3$ around the jet axis.  The missing
transverse energy $\met$ is defined as the magnitude of the vectorial
sum of the momenta of all lepton candidates and jets with
$\et>20~\GeV$ and $\abseta<5.0$.  Comparison between data and
simulation~\cite{JME-10-004,JME-10-005} shows good modelling of \met.

\begin{figure}
\centering
\includegraphics*[width=0.45\textwidth]{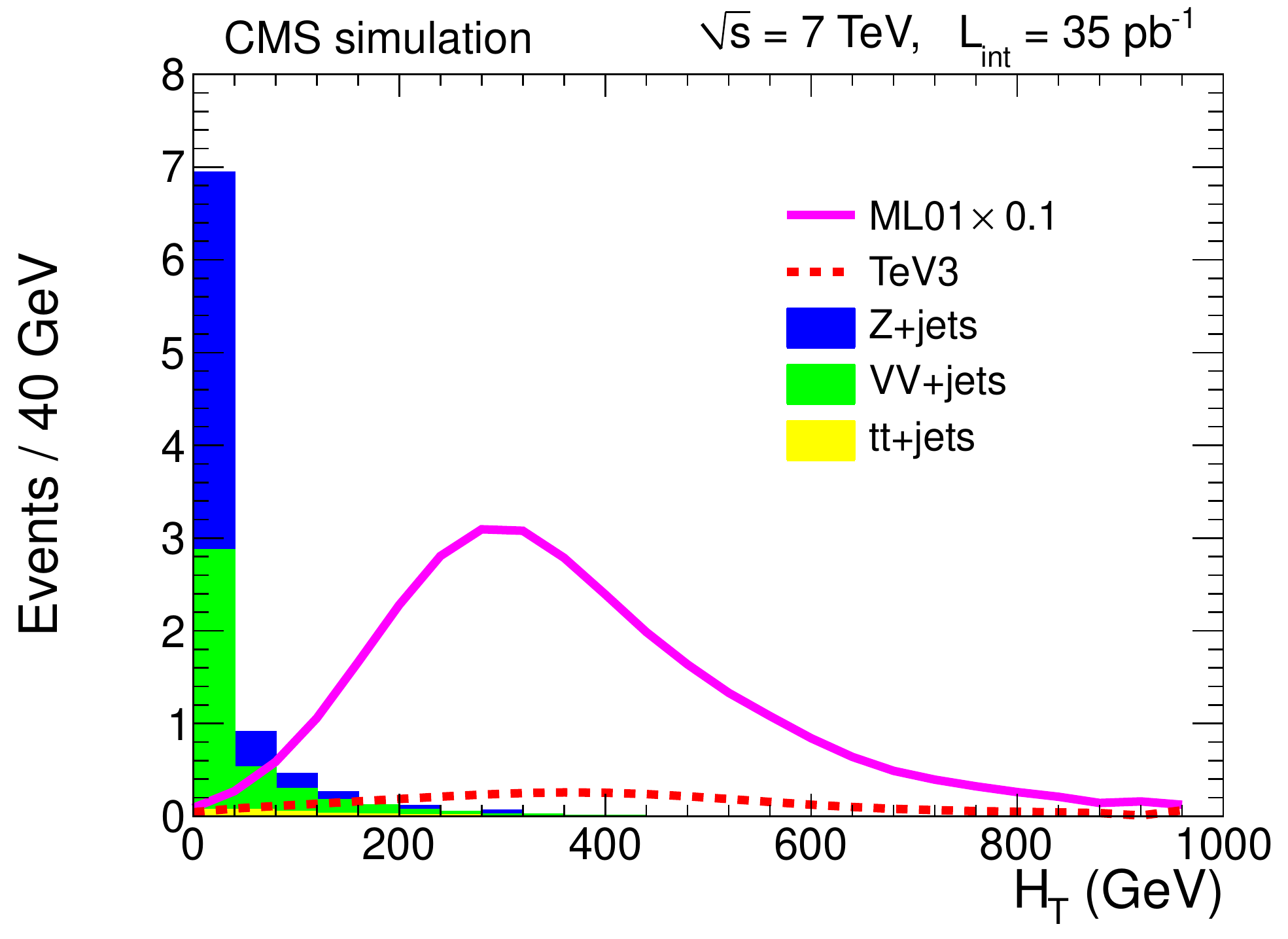}
\includegraphics*[width=0.45\textwidth]{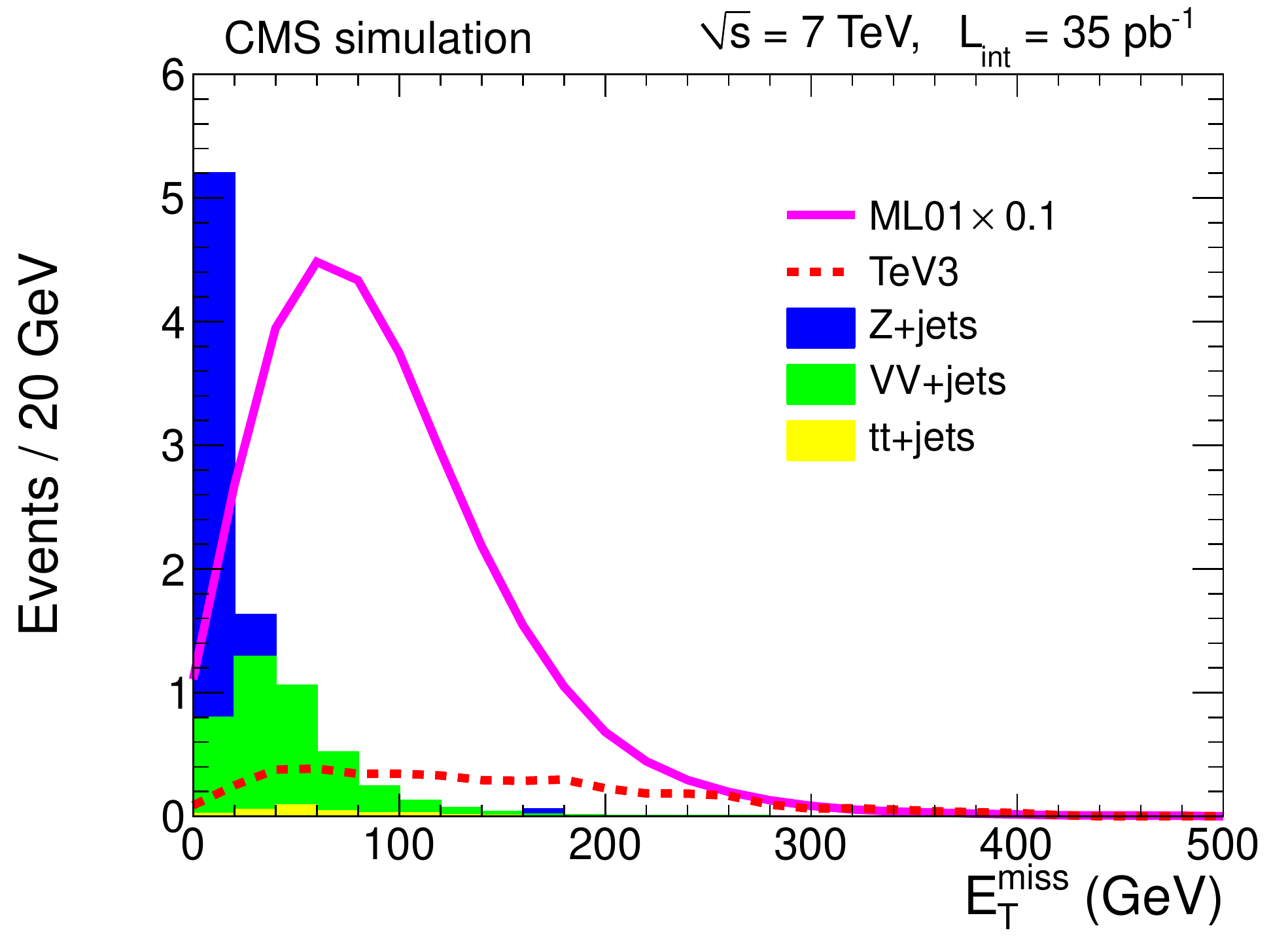}
\caption{The \jt\ (\xleft) and \met\ (\xright) distributions for SM background channels
(\textrm{Z}+jets, \ttbar, and \textrm{VV}+jets, where
$\textrm{V}=\textrm{W},\textrm{Z}$
and two SUSY
benchmark points for the simulation events that pass all
other requirements for the three-lepton events.
The ML01 and TeV3 benchmark points are defined in
Section~\ref{sec:intro} and details of the simulation are
given in Section~\ref{sec:background}.}

\label{fig:JT}
\end{figure}

\begin{figure}
\begin{center}
\includegraphics*[width=0.45\textwidth]{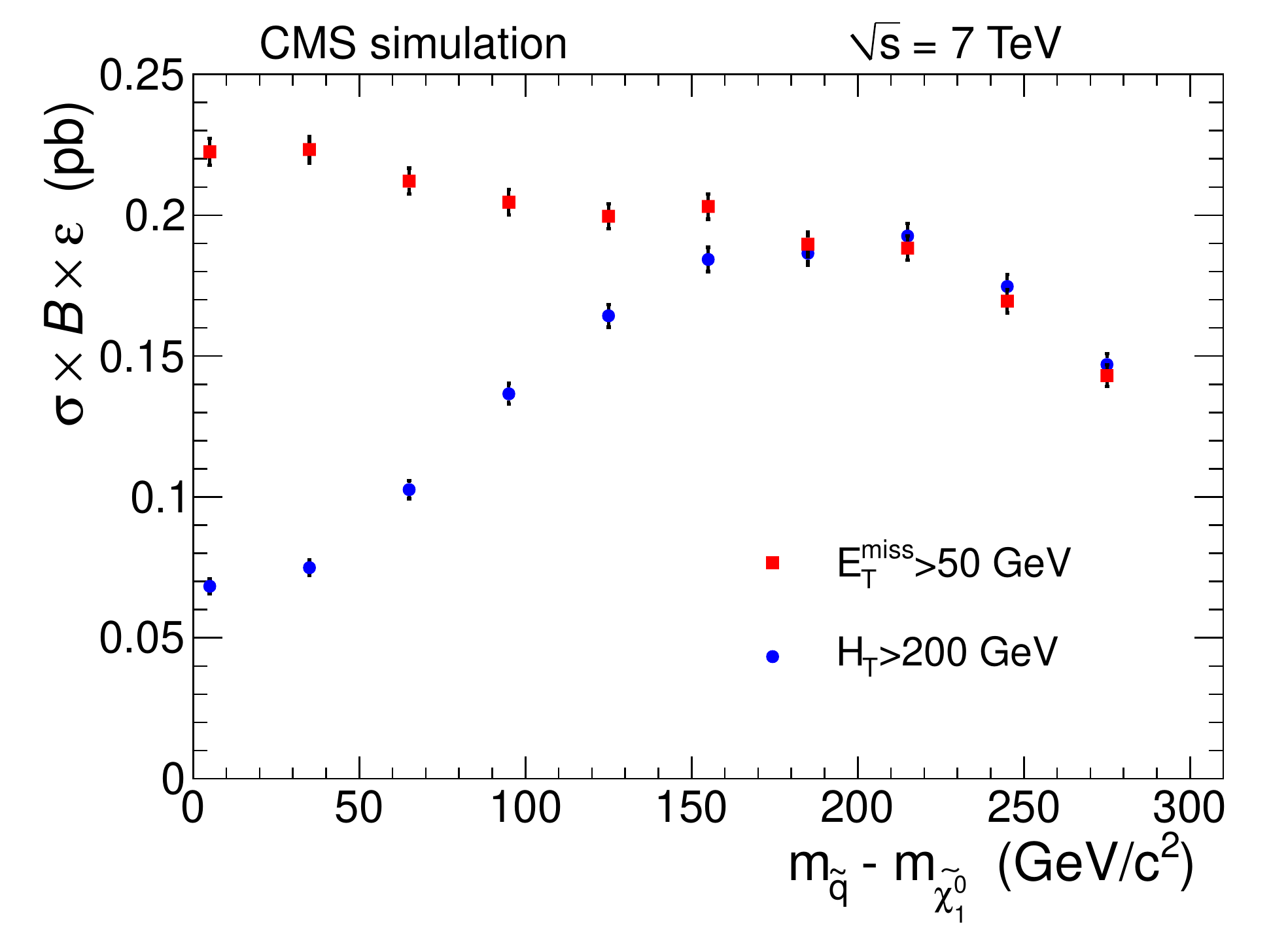}
\caption{
Effect of mass difference between squark and lightest neutralino
on cross section times branching fraction times efficiency
for an $\met> 50\GeV$ requirement (red squares) or for an $\jt >
200\GeV$ requirement (blue circles).  Less hadronic energy is released
if this difference becomes small, so the $\jt\ $ requirement loses
sensitivity in this region of parameter space.  The example is for
channels containing
two muons plus at least one electron or a tau. The slepton co-NLSP topology
used here is described in the text.
}
\label{fig:squarkbino}
\end{center}
\end{figure}

Both \jt\ and \met\ are good discriminating observables for
physics beyond the SM,
as demonstrated in Fig.~\ref{fig:JT}. In specific
 regions of parameter space one observable may be more effective than
 the other.  Figure~\ref{fig:JT} suggests that \jt\ has slightly superior discriminating
 power for the models we happen to consider here.  On the other hand,
\jt\ would be suppressed if the supersymmetric
production were dominated by electroweak processes,
as would be the case at the Tevatron~\cite{Aaltonen:2008pv}.
Another possibility is that the sparticle mass ordering
in the supersymmetric particle spectrum may result in reduced participation
of hadronic sparticles in the decay chain despite strong production,
resulting in negligible jet activity.
Figure~\ref{fig:squarkbino} illustrates this
 situation, showing the product of cross section, branching fraction, and
efficiency, i.e., event yield per unit integrated luminosity, as a function of the
 mass difference between the squark and lightest neutralino. The
slepton co-NLSP
supersymmetric topology illustrated here has degenerate
squarks with vanishing left-right mixing and right-handed sleptons
with masses of 500 and $185\GeVcc$, respectively, with a variable
lightest neutralino mass, and other superpartners decoupled.
The figure shows that the \jt requirement suppresses
sensitivity when neutralino and
squark masses are similar because squarks and gluinos fail
to participate in the decay chain, resulting in minimal hadronic
activity. By comparison,
\met is an appropriate discriminant in a multilepton search because
neutrino production generally accompanies \Pe , \Pgm , and \Pgt\
production.
Nonetheless, in order to retain search sensitivity beyond that of
dilepton searches, both \met\ and \jt\ selections should be used as sparingly
as possible.
The flexibility of the multichannel approach
allows us to selectively impose the \met\ or
\jt\ requirements in specific channels. Doing so maximises
sensitivity to new physics.

We exploit the background reduction ability of both \met\ and \jt\
as follows.
Events with $\met > 50\GeV$ ($\jt >200\GeV$) are said to satisfy
the \met\ ($H_T$) requirement. The justification for the values
chosen is evident from Fig.~\ref{fig:JT}.
 Another criterion for background
reduction is the ``\textrm{Z} veto'', in which the invariant mass of the
OS--SF lepton pairs is required to be outside the
75--105\GeVcc window.
A possible source of background is from the final state radiation in
$\textrm{Z}\rightarrow 2\ell (\ell=e,\mu)$
events undergoing a
$\gamma\rightarrow 2\ell$ conversion. Therefore, the \textrm{Z} veto requirement
is also applied to the
invariant mass $M(3\ell)$ of three leptons
for 3\Pe\ and \Pgm\Pgm\Pe\
events which have low \met\ and \jt.
As described below, these kinematic
selection criteria are applied together or separately as warranted
by the background level of the channel under consideration.

\subsection{Final kinematic selections}
\label{sec:finalkine}
In order to maximise sensitivity to diverse new physics scenarios, we
group the final selections into two broadly complementary domains.
As the name suggests, the {\it
hadronic} selection makes a uniform \jt\ requirement ($\jt >200\GeV$).
It reduces backgrounds to practically negligible values for channels
with electrons and muons. Both one- and
three-prong hadronic tau decays are reconstructed
using the PF technique.
For channels with OS--SF $\ell\ell$ pairs
plus $\tau$'s, the residual background from \textrm{Z}+jets is further reduced
with the \met\ requirement ($\met > 50\GeV$).  Only the \ttbar\
background then remains nonnegligible; about one event is expected
in \curlumi\ pb$^{-1}$ after the full selection.

The {\it inclusive} selection is based on the combined $\met > 50$ GeV
and Z-veto requirements for events with an OS--SF lepton pair.  These
events also must satisfy $M(2\ell)>12~\GeVcc$ to reject low mass Drell--Yan production
and the \PJgy\ and \PgU\ resonances.   In addition, candidate events are binned in exclusive channels characterised by total charge, number of lepton candidates, lepton flavours, high or low \met, and whether the \textrm{Z} veto described above is satisfied or not.
Isolated tracks are used to reconstruct the single-prong tau decays.

\section{Background Estimation}
\label{sec:background}

The main SM backgrounds in multilepton plus jet events originate
from \textrm{Z}+jets, double vector boson production
(\textrm{VV}+jets), \ttbar\ production, and QCD multijets.
Leptons associated with jets can be from heavy quark decays,
or with a lower probability, can be misidentified hadrons.
Leptons from heavy quark decays are suppressed
by the isolation requirement. The probability that a QCD event
includes three misidentified leptons is negligible.  Backgrounds from
cosmic rays are also found to be negligible.  Backgrounds from
beam-halo muons are included in the background estimate discussed
below.

The largest background remaining after the basic three-lepton
reconstruction originates from the \textrm{Z}+jets process, which in our
nomenclature includes the Drell--Yan process as well. The dileptons resulting
from these processes, together with misidentified isolated
tracks give rise to a trilepton background.
The probability that such
an isolated track is misidentified as a lepton is measured in control
samples where no signal should be present, such as in dijet samples. We
measure the probability for an isolated track to produce a
misidentified muon (electron) to be $2.2\pm0.6\%$ ($1.3^{+1.8}
_{-0.3} \%$).  The misidentification SM background for the three-lepton sample
is then obtained by multiplying the number of isolated tracks in the two-lepton
sample by this probability. In a similar way we estimate the
misidentified background for four-lepton events by examining
two-lepton events with two additional isolated tracks.  The large systematic
uncertainty on this rate is due to the difference in jet environment
in QCD and \textrm{Z}+jets control samples.  Such differences are expected
due to the variation of heavy quark content across the control
samples.

For channels with isolated tracks, we measure the SM background
by using the isolation sideband $0.2<I_{\mathrm{rel}}<1.0$ to extrapolate to
the signal region $I_{\mathrm{rel}}<0.15$.  In order to improve the statistical
error as well as to gain a systematic understanding of the
extrapolation process, we study the isolation distribution in various
QCD samples with different levels of jet activity and then evaluate the ratio
of events in the two isolation regions in the QCD sample that most
resembles the dilepton sample where the ratio is eventually
applied. The ratio of the numbers of isolated tracks in the two regions
is measured to be $15 \pm 3\%$.  The 3\% systematic uncertainty is
derived from the extent of variation of the ratio in these QCD control
samples. The ratio is then applied to the 2$\ell$ event sample. Because
the number of events after the \met\ selection is too small to be
useful, we derive the SM background in these channels by applying the
isolation probability ratio as well as the probability of a 2$\ell$
event to pass the \met\ selection to the full sample.

\begin{figure}
\begin{center}
\includegraphics*[width=0.45\textwidth]{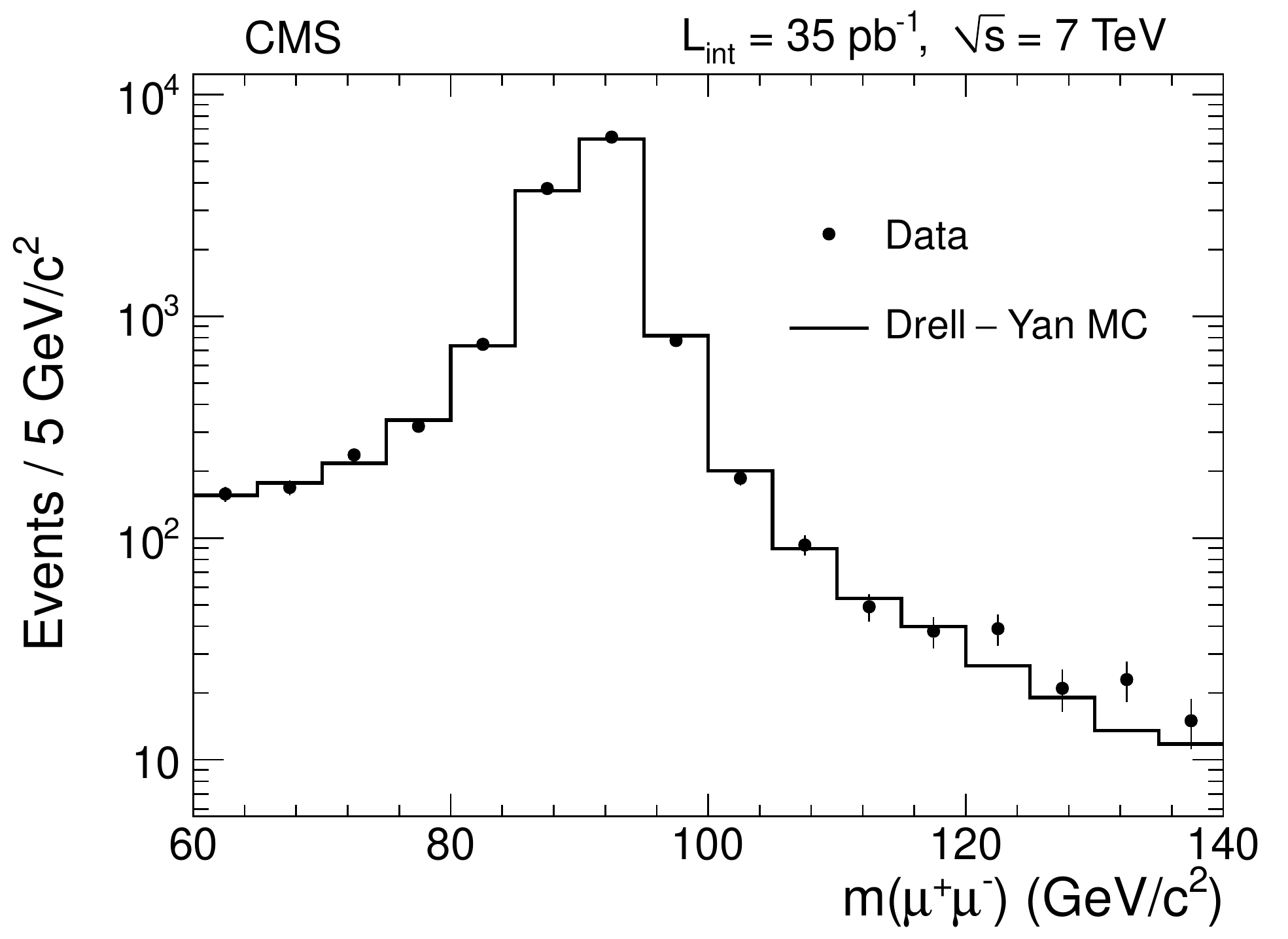}
\includegraphics*[width=0.45\textwidth]{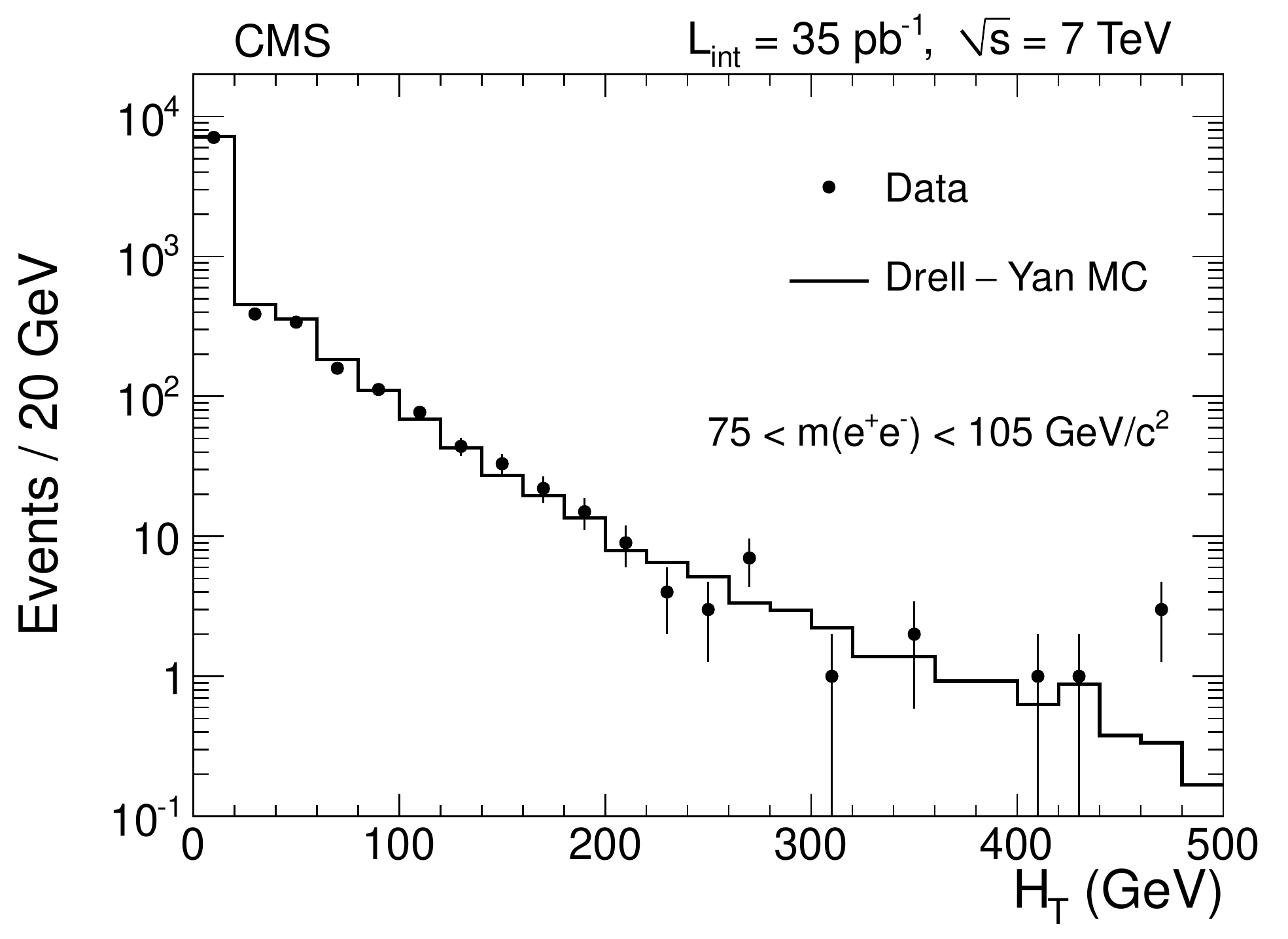}
\caption{
Two-lepton events in data, compared with the SM simulation. \xLeft: mass spectrum for \textrm{Z} $\to \mu\mu$.
\xRight: \jt\  distribution for \textrm{Z} $\to$ \Pe\Pe . Processes other than Drell--Yan are too rare to
be visible in these distributions.
}
\label{fig:massjt}

\end{center}
\end{figure}

Understanding of SM backgrounds at the three-lepton selection level as above is essential
before implementing the final kinematic selections.  We perform a detailed simulation of the detector response using {\sc Geant4}~\cite{Agostinelli:2002hh}
for \textrm{Z}/$\gamma^*$ + jets, \ttbar\ quark pairs, and double vector boson production
events generated using {\sc MadGraph}~\cite{Maltoni:2002qb}, and QCD events generated with
{\sc Pythia 8.1}~\cite{Sjostrand:2007gs}.  We use CTEQ6.6 parton distribution
functions~\cite{Nadolsky:2008zw}.  Already at  the dilepton level,
comparisons between data and simulation for distributions of the opposite-sign pair mass and
for \jt\ show good agreement for both muons and electrons. Figure~\ref{fig:massjt}
shows the mass spectrum for dimuon events and the \jt\ distribution for dielectron events.
After requiring a third lepton, the kinematic selections efficiently eliminate the \textrm{Z}+jets background.  The \ttbar\ and double vector boson backgrounds then come to the fore.

There is not sufficient data yet for a data-based estimate
of the  \ttbar\  background, so we use simulation, with the
contribution scaled to the measured \ttbar\ cross
section \cite{Khachatryan:2010ez}.
The \ttbar\ background comes primarily from
leptonic decays of both \textrm{W} bosons accompanied by a lepton from the \textrm{b} jets.
In order to verify the adequacy of simulation for
background estimation, we examine the
\Pe\Pgm\ dilepton distribution since
\ttbar\ contributes dominantly to it.
In particular, the spectrum of muons in this
sample which fail isolation requirements is
described well by the simulation whether
the muon originated promptly or not.  The same
is true of nonisolated tracks.
Agreement of these
distributions with the simulation gives
confidence that the semileptonic branching fractions of the
\textrm{b} quark and semileptonic form factors are reproduced
correctly by the simulation.
The \textrm{VV}+jets channels include
the irreducible background from \textrm{WZ}+jets with both vector bosons
decaying leptonically and the neutrino yielding missing energy, as
well as from \textrm{ZZ}+jets.  The simulation is used as these processes do
include prompt leptons, which are reasonably well described by
simulation~\cite{Khachatryan:2010xn}.

The lepton charge misidentification probability is generally less than
1\% for the lepton momenta typical for this search. The data-based
background estimation techniques described above automatically account
for charge misidentification background associated with
the \textrm{Z}+jets processes because the dilepton data sample used for
multilepton background estimation contains events with charge
misidentification. The probability of acquiring \textrm{WZ}
trilepton events with total charge of three units because of charge
misidentification is too small for the quantity of data considered here.

As a cross-check, the SM background events are  binned  in the two-dimensional isolation versus impact parameter plane. The background in the signal region, characterised by small isolation ($I_{\mathrm{rel}}<0.15$) and impact parameters ($d_{xy}<0.02$ cm),
is extrapolated from the three outside regions (''sidebands'') in this two-dimensional plot by assuming
the two variables $I_{\mathrm{rel}}$ and $d_{xy}$ to be
 uncorrelated, so both can be independently extrapolated.
This cross-check technique presently suffers from large statistical uncertainties, but the
resulting background estimates are consistent with those described above.

In summary, the  nonprompt backgrounds from \textrm{Z}+jets are measured from data, and the
methods described above
successfully predict the number of events in data samples dominated by SM processes.
The irreducible/prompt backgrounds from \ttbar\ and \textrm{VV}+jets are then obtained from simulation with high confidence.

\begin{table*}
\begin{center}
\footnotesize
\caption{
Summary of numbers of events in the various search channels (rows).
Channels with electrons and muons have been combined as $\ell\ell$,
with $\ell=$ \Pe\ or \Pgm , or $\ell\ell'$, if the flavours are
different. For the $\ell\ell\ell\ell$ channels different flavour
combinations are implied.
For the inclusive selection (upper table) isolated tracks are used
as proxy for the hadronic tau decays ($T$ channels), while for the
hadronic selection (lower table) PF tau reconstruction
($\tau$ channels) is used.
The rows for inclusive selection are aggregations
of selected subsets of channels used in the search.
The first three columns give the expected SM background events for the
dominant backgrounds after requiring the corresponding number of
leptons for each channel. The comparison with data at this stage is
given in the next two columns. The SM backgrounds are further reduced
using either inclusive or hadronic selection (see text)
and compared with data and signal expectations from the ML01 benchmark
point in the last columns.  Uncertainties are a combination of statistics plus
systematics relevant for SM background expectations.
\label{tab:bigtable}
}
\setlength{\extrarowheight}{1pt}
\vspace*{4mm}
\begin{tabular}{|l|ccc||cc||ccc|}
\cline{2-9}
\multicolumn{1}{c|}{}& \multicolumn{5}{c||}{After Lepton ID Requirements} & \multicolumn{3}{c|}{Inclusive Selection} \\
\hline
                 & \textrm{Z}+jets        &  \ttbar   & \textrm{VV}+jets  &$\Sigma$SM & Data  &$\Sigma$SM & Data &  ML01    \\
\hline
\multicolumn{1}{|c|}{Channel} & \multicolumn{8}{c|}{three-lepton channels}\rule{0pt}{4mm}\\
\hline
\textrm{OS}($\ell\ell$) $e$         & 1.7        & 0.1     &  1.2   &$4.4\pm1.5$& 6    &$0.1\pm0.1$&0 &121 \\
\textrm{OS}($\ell\ell$) $\mu $      & 2.8       & 0.2     &  1.7   &$4.7\pm0.5$& 6    &$0.1\pm0.1$&0  &  124    \\
\textrm{OS}($\ell\ell$) $T$          & 122      & 0.5     &  0.7   & $123\pm16$&127   &$0.4\pm0.1$&0  &   80          \\
$\ell\ell'T $            & 0.7       & 0.5     &  0.2   &$1.7\pm0.7$ &   3    &$0.4\pm0.2$&2
&  18.6         \\
\hline
\textrm{SS}($\ell\ell$) $ \ell'$        & 0.13       & 0.1     &  0.0   &$0.2\pm0.1$& 0  &$0.2\pm0.1$&0  &  2.8      \\
\textrm{SS}($\ell\ell$) $T$          & 0.25       & 0.0     &  0.1   &$0.7\pm0.4$& 3  &$0.1\pm0.1$&0  &   9.0  \\
$\ell TT $             & 47       & 0.3     &  0.1   &$48\pm9$   &   30   &$0.4 \pm 0.1  $   &   0
&   8.0       \\
\hline
\hline
 $\Sigma ~\ell\ell(\ell/T)$     & 127   & 1.4     &  3.8   &$135\pm16$ &  145   &$1.3\pm0.2$& 2
 &   356     \\
\hline\hline
\multicolumn{1}{|c|}{Channel} & \multicolumn{8}{c|}{four-lepton channels}\rule{0pt}{4mm}\\
\hline
$\ell\ell\ell\ell$             & 0          & 0        &  0.2   &$0.2\pm0.1$& 2   & 0         &  0   &  164   \\
$\ell\ell\ell T $           & 0          & 0        &  0.1   &$0.1\pm0.1$& 0   & 0         &  0   &    62    \\
$\ell\ell  TT $          & 0          & 0        &  0      &$0.0\pm0.1$& 0   & 0         &  0   &
21      \\
\hline\hline
$\Sigma~ \ell\ell(\ell/T)(\ell/T)$  & 0          & 0        &  0.3   &$0.3\pm0.1$& 2   & 0         & 0    &    247       \\
\hline \hline
\end{tabular}

\vspace*{4mm}
\begin{tabular}{|l|ccc||cc||ccc|}
\cline{2-9}
\multicolumn{1}{c|}{}& \multicolumn{5}{c||}{After Lepton ID Requirements} & \multicolumn{3}{c|}{Hadronic Selection}   \\
\hline
                 & \textrm{Z}+jets        &  \ttbar   & \textrm{VV}+jets  &$\Sigma$SM & Data  &$\Sigma$SM & Data &  ML01    \\
\hline
\multicolumn{1}{|c|}{Channel} & \multicolumn{8}{c|}{three-lepton channels}\\
\hline
\textrm{OS}($\ell\ell$) $e$         & 1.7        & 0.1     &  1.2   &$4.4\pm1.5$& 6    &$0.2\pm0.1$& 1
&142    \\
\textrm{OS}($\ell\ell$) $\mu $      & 2.8       & 0.2     &  1.7   &$4.7\pm0.5$& 6  &$0.1\pm0.1$& 0
&  121    \\
\textrm{OS}($\ell\ell$) $\tau$      & 476       & 2.7      &  3.9   &$484\pm77$ & 442  &$0.6\pm0.2$&1
&  68       \\
$\ell\ell'\tau $        &  4.7       &  2.9    &  0.6    &$11.2\pm2.5$&  10 &$0.4\pm0.1$& 1
&  12.3     \\
\hline
\textrm{SS}($\ell\ell$) $\ell'$        & 0.13       & 0.1     &  0.0   &$0.2\pm0.1$& 0 &$0.1\pm0.1$& 0
&  2.8      \\
\textrm{SS}($\ell\ell$) $\tau $     &  1.4       &  0.0    &   0.1  &$3.0\pm1.1$&  3&$0.0\pm0.1$&  0
&  6.9      \\
\hline
\hline
$\Sigma ~\ell\ell(\ell/\tau)$  &  487  &  6.0    &  7.5   &$507\pm77$ & 467  & $1.3\pm0.3$&3
&  350    \\
\hline
\hline
\multicolumn{1}{|c|}{Channel} & \multicolumn{8}{c|}{four-lepton channels}\\
\hline
$\ell\ell\ell\ell$             & 0          & 0        &  0.2   &$0.2\pm0.1$& 2    & 0         &  0    &  149    \\
$\ell\ell\ell\tau$          & 0          & 0        &  0.1    &$0.1\pm0.1$& 0   &  0         &  0    &  33     \\
$\ell\ell\tau\tau$       & 3.1        & 0.1     &  0.1   &$3.2\pm0.7$  & 5 &  0        &  0&  17     \\
\hline\hline
$\Sigma~ \ell\ell(\ell/\tau)(\ell/\tau)$ & 3.1       &  0.1    &   0.4  &$3.5\pm0.7$  & 5 &  0        &  0     &   199    \\
\hline \hline
\end{tabular}
\end{center}
\end{table*}

\section{Observations}

Table~\ref{tab:bigtable} shows the expected and observed numbers of three- and four-lepton events
in this search before and after the final  kinematic selections.
A tau candidate is indicated by $T$ for an isolated track as proxy for a hadronic tau decay
and $\tau$ for the PF tau selection. Channels containing OS--SF lepton
pairs are listed separately because they suffer from a larger
SM background expectation.
 The main SM backgrounds are given in the first three columns followed by the total SM background, which can be
 slightly larger than the sum of the previous columns, since it includes less significant backgrounds such as those involving initial and final state radiation.
Columns for the inclusive and hadronic kinematic selections show the number of events surviving all requirements.  
For the inclusive selection only the signal channels are shown, which require $\met > 50$ GeV, Z-veto,
and $M(2\ell)>12~\GeVcc$ for events with an OS--SF pair as discussed in Section~\ref{sec:finalkine}.
The control channels used in the limit setting are discussed in Section~\ref{sec:limitsetting}.
For the hadronic selection the background reduction comes from the $\jt>200$ GeV requirement.
The  sum of the SM backgrounds, mainly from \ttbar\ and the irreducible \textrm{VV}+jets backgrounds, is given as well.

Table~\ref{tab:bigtable} also shows signal expectations
for the slepton co-NLSP  benchmark point ML01 described earlier.
All cross sections for the benchmark point and those used in the following exclusion plots include  next-to-leading-order corrections calculated using {\sc Prospino}~\cite{Beenakker:1996ed}, which yields K factors in the
range 1.3--1.5.

Observations and SM expectations agree reasonably well.
We observe five three-lepton events worth noting.
An
\Pep\Pem\Pgt$^+$ event with $\jt = 246\GeV$
satisfies both the $\jt > 200\GeV$ and $\MET >50\GeV$ requirements.
So does an
\Pep\Pgmp\Pgt$^+$ event with $\jt = 384\GeV$. A
\Pgmp\Pgmm\Pep\ event satisfies the $\jt > 200\GeV$
requirement but not \met.
Two  \Pep\Pgmm$T^-$
events with \met\ of 70 and 101\GeV
fail the \jt\ requirement.   The largest background in the \Pe\Pgm$T$
channel is expected from \ttbar\ events,
and one event is indeed selected
as a \ttbar\ event in the CMS top selection~\cite{Khachatryan:2010ez}, but the other
fails the lepton $\pt$ requirement.

The four-lepton $\ell\ell\ell\ell$ row in
Table~\ref{tab:bigtable} also merits discussion since we
observe two \Pgmp\Pgmm\Pgmp\Pgmm\ events despite the SM expectation of
only 0.21 events.
One of the
events is completely consistent with the
$\textrm{ZZ} \rightarrow$ \Pgmp\Pgmm\Pgmp\Pgmm\
hypothesis.  The \textrm{ZZ}
invariant mass for this event is $212\GeVcc$ and it has
negligible \met. The second
event is unlikely to be a \textrm{ZZ} event, but it contains a \Pgmp\Pgmm\
pair with an invariant mass of
$80\GeVcc$ which is too close to the
\textrm{Z} mass
to pass the \textrm{Z} veto criterion.
Both events have small $\MET$, leptons
originating from the same vertex, and
minimal other activity.
\begin{figure}
\begin{center}
\includegraphics*[width=0.45\textwidth]{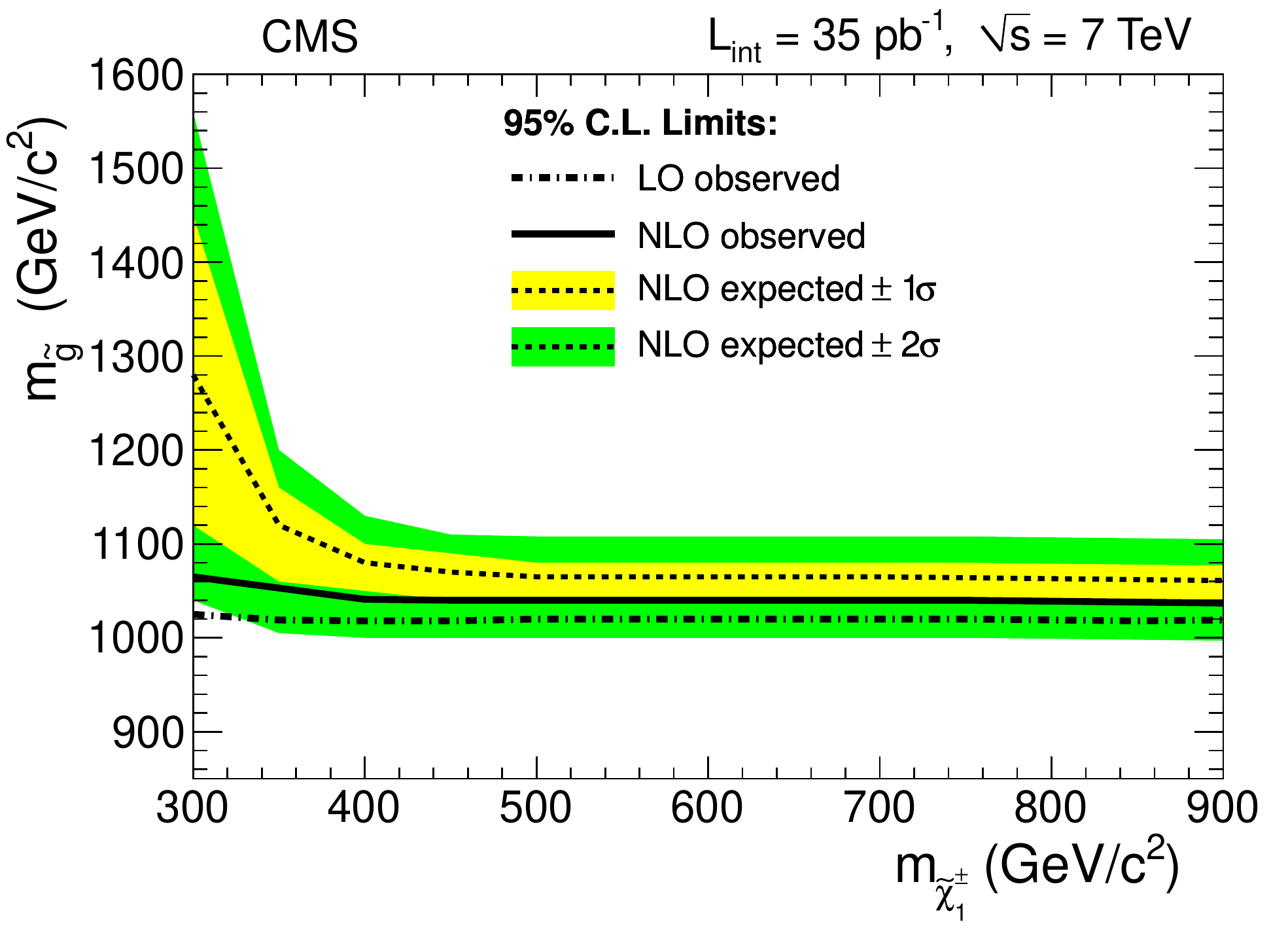}
\includegraphics*[width=0.45\textwidth]{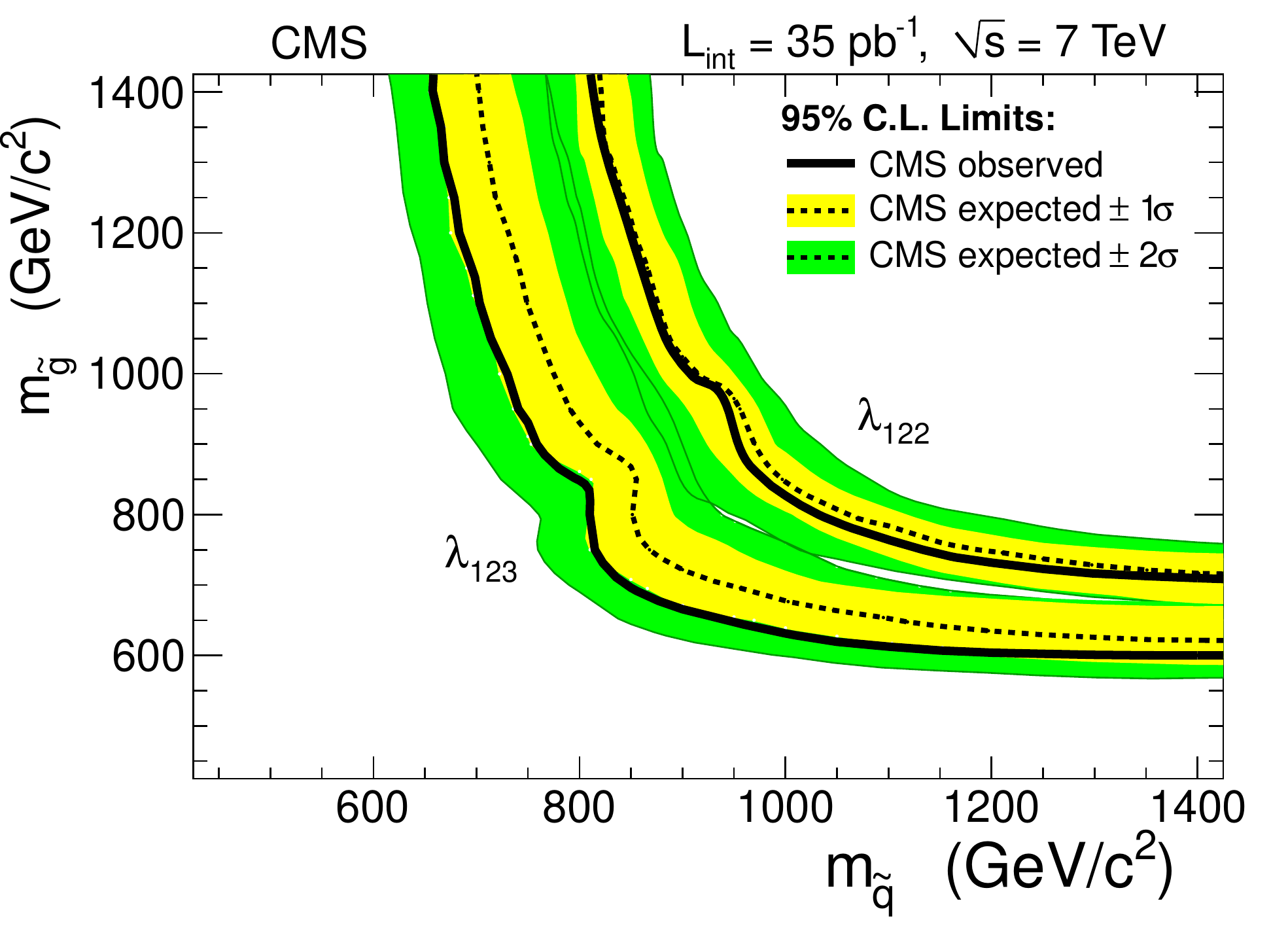}
\caption{
\xLeft: Limits on the slepton co-NLSP model as a function of the gluino
and wino-like chargino masses obtained by comparing with leading order (LO)
or next to leading order (NLO) cross sections.
\xRight: Limits for the R-parity violating scenario as a function of the gluino and degenerate squark
masses with either $\lambda_{122} \ne 0$ or $\lambda_{123} \ne 0$.
For both exclusions,
squark and slepton universality is enforced with
vanishing left-right mixing; mass relationships for other
superpartner masses are described in the text.
}
\label{fig:limit1}
\end{center}
\end{figure}

\subsection{Systematic uncertainties and statistical procedures}
\label{sec:limitsetting}
We discuss the sources of systematic
uncertainty and how they impact the search sensitivity before extracting
upper limits on the contributions from physics outside the SM.
All channels share systematic uncertainties for luminosity (11\%),
renormalization scales (10\%), parton distribution functions ($\leq
14\%$), and trigger efficiency ($\sim 5\%$). (Note that the luminosity
uncertainty subsequently decreased to 4\%, but the improvement
does not have significant implications for this result.)
The precision of lepton
selection efficiencies increases with lepton \pt. For a typical
slepton co-NLSP signal scenario which has leptons with \pt\ in excess
of 20\GeVc, the lepton identification and isolation efficiency
systematic uncertainty is $\sim1.5\%$ per lepton for muons and
electrons, as well as for isolated tracks.  However, CMSSM signals
result in lower \pt\ leptons, leading to a higher systematic
uncertainty on efficiency of $\sim3\%$ per lepton for muons and for
isolated tracks.  For low-energy electrons the systematic uncertainty
on the isolation efficiency can be as large as $\sim10\%$ because of
effects of synchrotron radiation in the high CMS solenoidal magnetic
field.  The uncertainty on the efficiency of PF tau
identification is studied using a comparison of
$\textrm{Z} \to $\Pgt\Pgt\ events in data and simulation.  For this study, events with a muon
plus hadronic tau decay are analysed, yielding a 30\%
systematic uncertainty~\cite{Maruyama:1355424}.

The impact of uncertainty from the jet energy scale for the \jt\
selection is $\leq 14\%$ as determined by varying the \jt\ requirement by $\pm 5\%$.
The jet-energy scale uncertainty~\cite{JME-10-010} has a small effect on the signal, since
the signal efficiency is high given the jet energy requirements; it varies in the range of 2--4\%, where the larger number is for the tau modes.

SM backgrounds derived from data suffer from large systematic
uncertainties because of the limited quantity of data in hand;
uncertainties on the misidentification rates are 30\% for the
PF taus, 20\% for tracks, $\sim 30\%$ for muons, and $\sim
80\%$ for electrons. These uncertainties are derived from extensive studies
in which misidentification rates are factorised into contributing
components such as isolation efficiency and the factorised pieces
are studied in different data sets.
Although these uncertainties appear to be large,
they do not affect the results significantly as the backgrounds are
small. The uncertainties on backgrounds derived from simulation are
dominated by the $\sim30\%$ uncertainty on the measured SM cross sections.

We utilise the agreement between the expected SM backgrounds and observations shown in
Table~\ref{tab:bigtable} to constrain new physics scenarios.
While being complementary in their approach, the two kinematic
selections overlap substantially.  This overlap must be removed in the
combination of the two selections to evaluate the search sensitivity for new
physics. For this purpose, we retain all events from the inclusive selection that satisfy
the additional requirement of $H_T<200$ GeV and all events from the
hadronic selection, which have by definition $H_T>200$ GeV.

There are 55 channels in the combination used for limit setting.
We include both $H_T>200$ GeV and $H_T<200$ GeV versions of the following 24 three- and
four-lepton channels: 2~OS$(\ell\ell)\Pe$; 2~OS$(\ell\ell)\mu$; 2~OS$(\ell\ell)\tau$; 2 $\ell\ell'\tau$; 
2~SS($\ell\ell)\ell'$; 2~SS($\ell\ell)\tau$; 5~$\ell\ell\ell\ell$; 4~$\ell\ell \ell \tau$; and 3~$\ell\ell\tau\tau$, 
where $\tau$ refers to either tau algorithm.
For the remaining seven channels, three require $H_T<200$ GeV and more than four leptons, with up to two taus, and four require three leptons with two taus:
$\Sigma Q=\pm3$; $\met > 50$ GeV and $H_T > 200$ GeV; $\met > 50$ GeV and on-Z;
and $\met < 50$ GeV and on-Z.

The statistical model uses a Poisson distribution for the number of events in each channel, while the nuisance parameters are modeled with a Gaussian, truncated to be always positive.
The significant nuisance parameters are the luminosity uncertainty, trigger efficiency, and lepton identification
efficiencies. The expected value in the model is the sum of the signal and the expected backgrounds.
We set 95\% confidence level (CL)
upper limits on the signal parameters and cross sections using a Bayesian method with a flat prior.
We check the stability of the result with respect to nuisance
constraints selection by substituting log-normal constraints for the Gaussian ones, and find
the upper limit results to be stable within 3\%.
The statistical model is implemented in the program package {\sc RooStats}~\cite{roostats}.
We apply these upper limits on the contribution of new physics for the following SUSY scenarios.

\subsection{Slepton co-NLSP}
In supersymmetry, multilepton final states arise naturally in the
subset of GMSM parameter space where the right-handed sleptons are
flavour-degenerate and at the bottom of the Minimal Supersymmetric
Standard Model (MSSM) mass spectrum.  The Higgsinos are decoupled.
Supersymmetric production
proceeds mainly through pairs of squarks and/or gluinos.  Cascade
decays of these states eventually pass sequentially through the
lightest neutralino ($\tilde{g},\tilde{q}\rightarrow \chi^0 + X$),
which decays into a slepton and a lepton
($\chi^0 \rightarrow \tilde{\ell}^\pm \ell^{\mp}$). Each of the
essentially degenerate right-handed sleptons promptly decays to the
Goldstino component of the almost massless and non-interacting
gravitino and a lepton ($\tilde{\ell} \to \tilde{G}\ell$) thus
yielding events with four or more hard leptons and missing energy.
Such scenarios have a high cross section with little
background~\cite{Alves:2011wf}.

The 95\% CL exclusion limits for the slepton co-NLSP model is
 shown in the \xleft panel of Fig.~\ref{fig:limit1}.  Deviation from
 the expected limit is due to a modest data excess.  The result corresponds
 to a limit of $\approx$ 6 events on the signal yield, and
a slepton co-NLSP benchmark 95\% CL upper limit on the
 cross section of $\sigma_{95}=0.2$--$0.4\unit{pb}$. Squark and gluino masses
 of up to $830\GeVcc$ and $1040\GeVcc$ are excluded.

\subsection{R-parity violation}
Although R-parity is often
assumed to be conserved, the most general formulation of the MSSM
superpotential contains R-parity violating  couplings $\lambda_{ijk}$,
where $i,j,$ and $k$ are generation indices.
We  study models in which lepton-number-violating decays are allowed, but  baryon number is conserved, so these models are  not constrained by limits on proton lifetime which require both $B$ and $L$ violation.

Events with four or more charged leptons
in the final state originate from the production of pairs of squarks or
gluinos, each of which cascade decays down to the LSP, which in the model
considered here is the  neutralino. Each
neutralino decays to two charged leptons and a neutrino.
Any nonzero value of
$\lambda_{ijk}$ causes the  neutralino to decay, yielding
multilepton final states. The actual value of $\lambda_{ijk}$ simply
determines the lifetime and hence the decay length of the neutralino.
We consider $\lambda_{ijk}$ to be sufficiently large so that the decay is prompt, the
exclusion limits are independent of $\lambda_{ijk}$ value, and thus the search
is sensitive only to the sparticle masses. We consider the cases of
nonzero $\lambda_{122}$ and nonzero $\lambda_{123}$ separately. For the
$\lambda_{122}$ coupling, the two charged leptons in each neutralino decay are electron and/or muon, while for $\lambda_{123}$, one of the charged leptons is a tau, and the other an electron or muon~\cite{dmitry_hits_thesis}.

The 95\%  exclusion limits in the squark-gluino mass plane
obtained using the inclusive kinematic selection
are shown in the \xright panel
of  Fig.~\ref{fig:limit1} for a topology with fixed $m_{\chi^0_1}=300\GeVcc$,
$m_{\tilde{\ell}_L} = m_{\tilde{\ell}_R}= 1000\GeVcc$, and with the wino
and the Higgsino decoupled.
The bumps in the contour plot are due to the fact that when the squark mass
is larger than the gluino mass there are two additional jets in the event.
This lowers the efficiency of the lepton isolation requirement and therefore
decreases the signal acceptance.
 The limits for the
$\lambda_{123}$ coupling are lower because of the lower acceptance for taus.
These results  substantially extend previous exclusion limits from
CDF and D0 based on integrated luminosities of 350 pb$^{-1}$~\cite{Abazov:2006nw,Abulencia:2007mp}.

\begin{figure}
\begin{center}
\includegraphics*[width=0.45\textwidth]{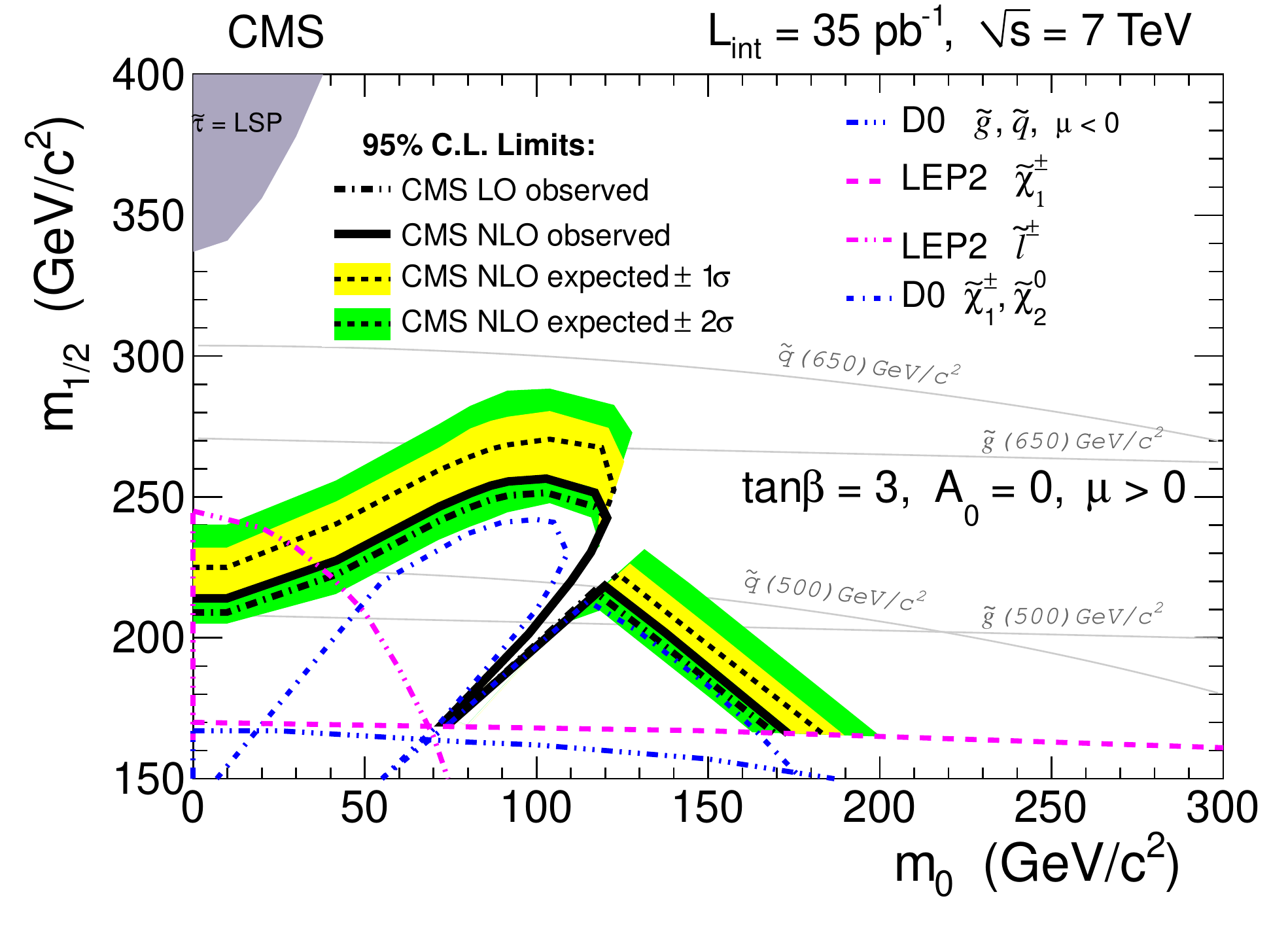}
\includegraphics*[width=0.45\textwidth]{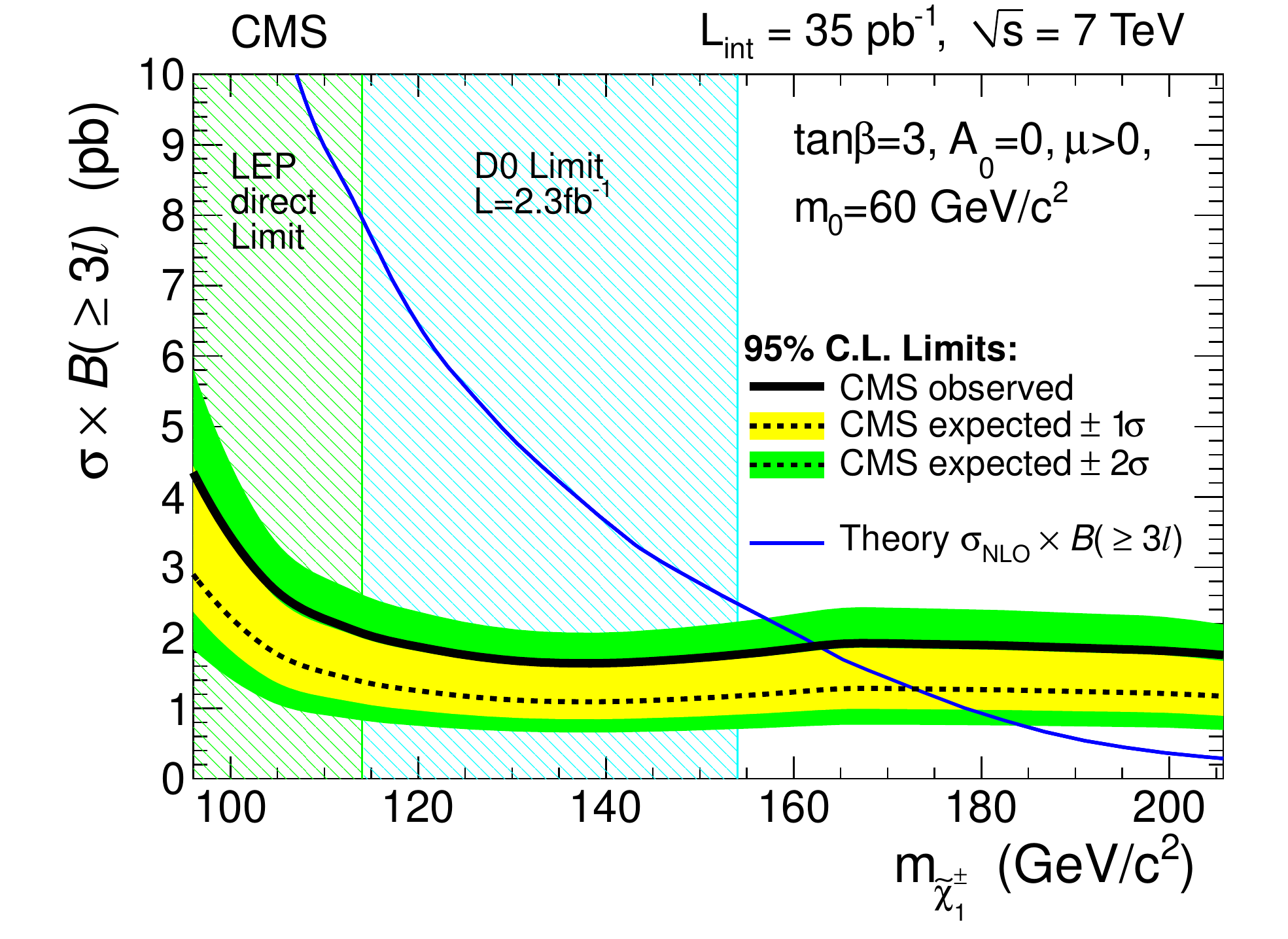}
\caption{\xLeft: excluded region  for the mSUGRA/CMSSM scenario  along with the limits from the multilepton searches from the Tevatron~\cite{Abazov:2009zi} and the exclusion derived from  slepton and chargino limits from LEP~\cite{lepsusy,Heister:2001nk,Heister:2003zk,Abdallah:2003xe,Achard:2003ge,Abbiendi:2003ji}.
The region below the lines is excluded at 95\% CL.
\xRight: the expected and observed upper limits on the cross section times branching  ratio $\sigma\times B(3\ell)$ as a function of the chargino mass.
The  theoretical curve crosses the observed 95\% CL upper limit on the cross section at $163\GeVcc$, thus excluding charginos below this mass for the values of $m_0$, $\azero$, and $\tb$ indicated
in the figure.
 For comparison the regions excluded by LEP (from slepton limits~\cite{lepsusy,Heister:2001nk,Heister:2003zk,Abdallah:2003xe,Achard:2003ge,Abbiendi:2003ji}),
Tevatron chargino-neutralino production~\cite{Abazov:2009zi}, and
Tevatron squark-gluino production~\cite{D0sqgl}
are indicated as well. This and other results have the other MSSM parameters fixed at
 $\tan\beta=3$, $\azero=0$, and $\mu>0$ except~\cite{D0sqgl}, which uses $\mu<0$.
 \label{fig:limit3} }
\end{center}
\end{figure}

\subsection{mSUGRA/CMSSM scenario}
For the mSUGRA/CMSSM~\cite{MSUGRA,CMSSM} scenario,
limits in the $m_0$-$m_{1/2}$ plane are shown in Fig.~\ref{fig:limit3} for
$\azero=0,\tb=3,$ and $\mu>0$. The TeV3 benchmark point defined above is close to the excluded limit from the Tevatron data; the total number of expected events after all cuts is $\approx$ 7 for the \curlumi\ \pbinv\ data sample.  As can be seen, our results extend the excluded region in comparison with previous results from LEP and the Tevatron.
For small values of $m_0$ the sleptons  can become lighter than the gauginos, so the gauginos will decay into slepton and lepton (two-body decay), although for larger values of $m_0$  three-body decays will dominate.
While for two-body decays the branching fraction into leptons is 100\%, it decreases rapidly for three-body decays.
In the transition region from two- to three-body decays the leptons become soft and fail the $\pt$ requirement~\cite{Aaltonen:2008pv}. Exclusion is therefore not possible, as shown by the non-excluded region between the two- and three-body decay regions.
We exclude gluino masses
up to $628\GeVcc$ for this choice of parameters.
The 95\% CL upper limit on the cross section times branching
fraction into 3$\ell$ varies from $\sigma_{95}=0.8$ to 2~pb. The
sensitivity to the chargino mass can be seen in the \xright
panel of Fig.~\ref{fig:limit3}, where the NLO cross section
for $m_0=60\GeVcc$ equals the 95\% CL experimental limit of $\sigma_{95}=2$ pb
for chargino mass of $163\GeVcc$. Therefore, chargino masses
above this value cannot be excluded.

\section{Conclusion}
We have performed a search for physics beyond the SM using multilepton
final states.  Taking advantage of the high centre-of-mass energy at the
LHC, we were able to probe new regions of the MSSM parameter space. Our search
complements those at the Tevatron, which are mostly sensitive
to electroweak gaugino production via quark-antiquark interaction,
while the result presented here is mostly sensitive to gluino and
squark production via quark-gluon or gluon-gluon interactions.

The results of this search are consistent with SM expectations.  In
the CMSSM parameter space, gluino masses up
to $628\GeVcc$ are thus excluded for specific SUSY parameters. This result is better than
the prior multilepton results from the Tevatron, but is in the region
already ruled out by other hadronic searches at the
LHC~\cite{Khachatryan:2011tk,PhysRevLett.106.131802}. However, the following two regions of MSSM
are not accessible to hadronic searches.  With
gravitinos as LSP and sleptons as co-NLSP, we are able to exclude
squark and gluino masses of up to $830\GeVcc$ and $1040\GeVcc$,
respectively.  We are also able to exclude models with
leptonic R-parity violation for gluino masses up to 600--700$\GeVcc$ depending on the choice
of parameters.
In both cases our search significantly extends into the regions of SUSY
parameter space not accessible to multilepton searches at the Tevatron.

\section*{Acknowledgements}
We thank Michael Park and Yue Zhao (Rutgers) for assistance in simulating theoretical models.

We wish to congratulate our colleagues in the CERN accelerator departments for the excellent performance of the LHC machine. We thank the technical and administrative staff at CERN and other CMS institutes, and acknowledge support from: FMSR (Austria); FNRS and FWO (Belgium); CNPq, CAPES, FAPERJ, and FAPESP (Brazil); MES (Bulgaria); CERN; CAS, MoST, and NSFC (China); COLCIENCIAS (Colombia); MSES (Croatia); RPF (Cyprus); Academy of Sciences and NICPB (Estonia); Academy of Finland, MEC, and HIP (Finland); CEA and CNRS/IN2P3 (France); BMBF, DFG, and HGF (Germany); GSRT (Greece); OTKA and NKTH (Hungary); DAE and DST (India); IPM (Iran); SFI (Ireland); INFN (Italy); NRF and WCU (Korea); LAS (Lithuania); CINVESTAV, CONACYT, SEP, and UASLP-FAI (Mexico); MSI (New Zealand); PAEC (Pakistan); SCSR (Poland); FCT (Portugal); JINR (Armenia, Belarus, Georgia, Ukraine, Uzbekistan); MST and MAE (Russia); MSTD (Serbia); MICINN and CPAN (Spain); Swiss Funding Agencies (Switzerland); NSC (Taipei); TUBITAK and TAEK (Turkey); STFC (United Kingdom); DOE and NSF (USA).

Individuals have received support from the Marie-Curie programme and the European Research Council (European Union); the Leventis Foundation; the A. P. Sloan Foundation; the Alexander von Humboldt Foundation; the Associazione per lo Sviluppo Scientifico e Tecnologico del Piemonte (Italy); the Belgian Federal Science Policy Office; the Fonds pour la Formation \`a la Recherche dans l'Industrie et dans l'Agriculture (FRIA-Belgium); and the Agentschap voor Innovatie door Wetenschap en Technologie (IWT-Belgium).

\bibliography{auto_generated}   
\cleardoublepage \appendix\section{The CMS Collaboration \label{app:collab}}\begin{sloppypar}\hyphenpenalty=5000\widowpenalty=500\clubpenalty=5000\input{SUS-10-008-authorlist.tex}\end{sloppypar}
\end{document}

%% file: SUS-10-008-authorlist.tex
\textbf{Yerevan Physics Institute,  Yerevan,  Armenia}\\*[0pt]
S.~Chatrchyan, V.~Khachatryan, A.M.~Sirunyan, A.~Tumasyan
\vskip\cmsinstskip
\textbf{Institut f\"{u}r Hochenergiephysik der OeAW,  Wien,  Austria}\\*[0pt]
W.~Adam, T.~Bergauer, M.~Dragicevic, J.~Er\"{o}, C.~Fabjan, M.~Friedl, R.~Fr\"{u}hwirth, V.M.~Ghete, J.~Hammer\cmsAuthorMark{1}, S.~H\"{a}nsel, M.~Hoch, N.~H\"{o}rmann, J.~Hrubec, M.~Jeitler, W.~Kiesenhofer, M.~Krammer, D.~Liko, I.~Mikulec, M.~Pernicka, H.~Rohringer, R.~Sch\"{o}fbeck, J.~Strauss, A.~Taurok, F.~Teischinger, P.~Wagner, W.~Waltenberger, G.~Walzel, E.~Widl, C.-E.~Wulz
% ...
\vskip\cmsinstskip
% ...
1:~~Also at CERN, European Organization for Nuclear Research, Geneva, Switzerland\\
% ...


%% file: SUS-10-008_temp.bbl
\providecommand{\href}[2]{#2}\begingroup\raggedright\begin{thebibliography}{10}%
\makeatletter
\providecommand{\hrefCMSnoop }[0]{\@secondoftwo}%
\makeatother

\bibitem{Nilles:1983ge}
\hrefCMSnoop {} {H.~P. Nilles, ``Supersymmetry, Supergravity and Particle
  Physics'',} \textit{ Phys. Rept.} \textbf{ 110} (1984) 1.
\href{http://dx.doi.org/10.1016/0370-1573(84)90008-5}{\texttt{
  doi:10.1016/0370-1573(84)90008-5}}.

\bibitem{Haber:1984rc}
\hrefCMSnoop {} {H.~E. Haber and G.~L. Kane, ``The Search for Supersymmetry:
  Probing Physics Beyond the Standard Model'',} \textit{ Phys. Rept.} \textbf{
  117} (1985) 75.
\href{http://dx.doi.org/10.1016/0370-1573(85)90051-1}{\texttt{
  doi:10.1016/0370-1573(85)90051-1}}.

\bibitem{deBoer:1994dg}
\hrefCMSnoop {} {W.~de~Boer, ``Grand unified theories and supersymmetry in
  particle physics and cosmology'',} \textit{ Prog. Part. Nucl. Phys.} \textbf{
  33} (1994) 201.
\href{http://dx.doi.org/10.1016/0146-6410(94)90045-0}{\texttt{
  doi:10.1016/0146-6410(94)90045-0}}.

\bibitem{Khachatryan:2011tk}
\hrefCMSnoop {} {{ CMS} Collaboration, ``Search for Supersymmetry in pp
  Collisions at 7 {TeV} in Events with Jets and Missing Transverse Energy'',}
  \textit{ Phys. Lett.} \textbf{ B698} (2011) 196.
  \href{http://dx.doi.org/10.1016/j.physletb.2011.03.021}{\texttt{
  doi:10.1016/j.physletb.2011.03.021}}.

\bibitem{PhysRevLett.106.131802}
\hrefCMSnoop {} {{ ATLAS} Collaboration, ``Search for Supersymmetry Using Final
  States with One Lepton, Jets, and Missing Transverse Momentum with the
  {ATLAS} Detector in $\sqrt{s}$=7 {TeV} pp Collisions'',} \textit{ Phys. Rev.
  Lett.} \textbf{ 106} (2011) 131802.
  \href{http://dx.doi.org/10.1103/PhysRevLett.106.131802}{\texttt{
  doi:10.1103/PhysRevLett.106.131802}}.

\bibitem{Aaltonen:2008pv}
\hrefCMSnoop {} {{ CDF} Collaboration, ``Search for Supersymmetry in $p
  \bar{p}$ Collisions at $\sqrt{s}$ = 1.96~{TeV} Using the Trilepton Signature
  for Chargino-Neutralino Production'',} \textit{ Phys. Rev. Lett.} \textbf{
  101} (2008) 251801.
\href{http://dx.doi.org/10.1103/PhysRevLett.101.251801}{\texttt{
  doi:10.1103/PhysRevLett.101.251801}}.

\bibitem{Dube:2008kf}
\hrefCMSnoop {} {S.~Dube {et~al.}, ``Addressing the Multi-Channel Inverse
  Problem at High Energy Colliders: A Model Independent Approach to the Search
  for New Physics with Trileptons'',} (2008).
  \href{http://www.arXiv.org/abs/0808.1605}{\texttt{ arXiv:0808.1605}}.


\bibitem{Ruderman:2010kj}
\hrefCMSnoop {} {J.~T. Ruderman and D.~Shih, ``Slepton co-{NLSP}s at the
  {T}evatron'',} (2010). \href{http://www.arXiv.org/abs/1009.1665}{\texttt{
  arXiv:1009.1665}}.


\bibitem{Abazov:2009zi}
\hrefCMSnoop {} {{ D0} Collaboration, ``Search for associated production of
  charginos and neutralinos in the trilepton final state using 2.3 fb$^{-1}$ of
  data'',} \textit{ Phys. Lett.} \textbf{ B680} (2009) 34.
\href{http://dx.doi.org/10.1016/j.physletb.2009.08.011}{\texttt{
  doi:10.1016/j.physletb.2009.08.011}}.

\bibitem{Forrest:2009gm}
\hrefCMSnoop {} {{ CDF} Collaboration, ``Search for Supersymmetry in
  $p\overline{p}$ Collisions at $\sqrt{s} =1.96$ {TeV} Using the Trilepton
  Signature of Chargino-Neutralino Production'',} (2009).
  \href{http://www.arXiv.org/abs/0910.1931}{\texttt{ arXiv:0910.1931}}.


\bibitem{Abazov:2006nw}
\hrefCMSnoop {} {{ D0} Collaboration, ``Search for {R}-parity violating
  supersymmetry via the {$LL \overline{E}$} couplings $\lambda_{121}$,
  $\lambda_{122}$ or $\lambda_{133}$ in $p \bar{p}$ collisions at $\sqrt{s}$ =
  1.96~{TeV}'',} \textit{ Phys. Lett.} \textbf{ B638} (2006) 441.
\href{http://dx.doi.org/10.1016/j.physletb.2006.05.077}{\texttt{
  doi:10.1016/j.physletb.2006.05.077}}.

\bibitem{Abulencia:2007mp}
\hrefCMSnoop {} {{ CDF} Collaboration, ``Search for anomalous production of
  multilepton events in $p \bar{p}$ collisions at $\sqrt{s}$ = 1.96~{TeV}'',}
  \textit{ Phys. Rev. Lett.} \textbf{ 98} (2007) 131804.
\href{http://dx.doi.org/10.1103/PhysRevLett.98.131804}{\texttt{
  doi:10.1103/PhysRevLett.98.131804}}.

\bibitem{MSUGRA}
\hrefCMSnoop {} {A.~H. Chamseddine, R.~Arnowitt, and P.~Nath, ``Locally
  Supersymmetric Grand Unification'',} \textit{ Phys. Rev. Lett.} \textbf{ 49}
  (1982) 970. \href{http://dx.doi.org/10.1103/PhysRevLett.49.970}{\texttt{
  doi:10.1103/PhysRevLett.49.970}}.

\bibitem{CMSSM}
\hrefCMSnoop {} {G.~Kane {et~al.}, ``Study of constrained minimal
  supersymmetry'',} \textit{ Phys. Rev.} \textbf{ D49} (1994) 6173.
  \href{http://dx.doi.org/10.1103/PhysRevD.49.6173}{\texttt{
  doi:10.1103/PhysRevD.49.6173}}.

\bibitem{Dimopoulos:1996va}
\hrefCMSnoop {} {S.~Dimopoulos, S.~D. Thomas, and J.~D. Wells, ``Implications
  of low energy supersymmetry breaking at the {F}ermilab Tevatron'',} \textit{
  Phys. Rev.} \textbf{ D54} (1996) 3283.
\href{http://dx.doi.org/10.1103/PhysRevD.54.3283}{\texttt{
  doi:10.1103/PhysRevD.54.3283}}.

\bibitem{Culbertson:2000am}
\hrefCMSnoop {} {{ SUSY Working Group} Collaboration, ``Low-scale and
  gauge-mediated supersymmetry breaking at the {F}ermilab {T}evatron Run
  {II}'',} (2000). \href{http://www.arXiv.org/abs/hep-ph/0008070}{\texttt{
  arXiv:hep-ph/0008070}}.


\bibitem{Alves:2011wf}
\hrefCMSnoop {} {D.~Alves {et~al.}, ``Simplified Models for {LHC} New Physics
  Searches'',} (2011). \href{http://www.arXiv.org/abs/1105.2838}{\texttt{
  arXiv:1105.2838}}.


\bibitem{Chatrchyan:2009hb}
\hrefCMSnoop {} {{ CMS} Collaboration, ``Commissioning of the {CMS} Experiment
  and the Cosmic Run at Four {Tesla}'',} \textit{ JINST} \textbf{ 5} (2010)
  T03001.
\href{http://dx.doi.org/10.1088/1748-0221/5/03/T03001}{\texttt{
  doi:10.1088/1748-0221/5/03/T03001}}.

\bibitem{:2008zzk}
\hrefCMSnoop {} {{ CMS} Collaboration, ``The {CMS} experiment at the {CERN}
  {LHC}'',} \textit{ JINST} \textbf{ 3} (2008) S08004.
\href{http://dx.doi.org/10.1088/1748-0221/3/08/S08004}{\texttt{
  doi:10.1088/1748-0221/3/08/S08004}}.

\bibitem{EGM-10-004}
\href {http://cdsweb.cern.ch/record/1299116} {{ CMS} Collaboration, ``Electron
  Reconstruction and Identification at $\sqrt{s} = 7$ {TeV}'',} \textit{ CMS
  Physics Analysis Summary} \textbf{ CMS-PAS-EGM-10-004} (2010).

\bibitem{MUO-10-002}
\href {http://cdsweb.cern.ch/record/1279140} {{ CMS} Collaboration,
  ``Performance of muon identification in pp collisions at $\sqrt{s}$ = 7
  {TeV}'',} \textit{ CMS Physics Analysis Summary} \textbf{ CMS-PAS-MUO-10-002}
  (2010).

\bibitem{PFT-10-002}
\href {http://cdsweb.cern.ch/record/1279341} {{ CMS} Collaboration,
  ``Commissioning of the Particle-Flow Reconstruction in Minimum-Bias and Jet
  Events from {\Pp\Pp} Collisions at 7 {TeV}'',} \textit{ CMS Physics Analysis
  Summary} \textbf{ CMS-PAS-PFT-10-002} (2010).

\bibitem{PFT-10-004}
\href {http://cdsweb.cern.ch/record/1279358} {{ CMS} Collaboration, ``Study of
  tau reconstruction algorithms using {\Pp\Pp} collisions data collected at
  $\sqrt{s} = 7\,${TeV} with {CMS} detector at {LHC}'',} \textit{ CMS Physics
  Analysis Summary} \textbf{ CMS-PAS-PFT-10-004} (2010).

\bibitem{PFT-08-001}
\href {http://cdsweb.cern.ch/record/1198228} {{ CMS} Collaboration, ``CMS
  Strategies for tau reconstruction and identification using particle-flow
  techniques'',} \textit{ CMS Physics Analysis Summary} \textbf{
  CMS-PAS-PFT-08-001} (2009).

\bibitem{JME-10-004}
\href {http://cdsweb.cern.ch/record/1279142} {{ CMS} Collaboration, ``Missing
  Transverse Energy Performance in Minimum-Bias and Jet Events from
  Proton-Proton Collisions at $\sqrt{s}=7$ {TeV}'',} \textit{ CMS Physics
  Analysis Summary} \textbf{ CMS-PAS-JME-10-004} (2010).

\bibitem{JME-10-005}
\href {http://cdsweb.cern.ch/record/1294501} {{ CMS} Collaboration, ``{CMS}
  {MET} Performance in Events Containing Electroweak Bosons from pp Collisions
  at $\sqrt{s}=7$ {TeV}'',} \textit{ CMS Physics Analysis Summary} \textbf{
  CMS-PAS-JME-10-005} (2010).

\bibitem{Agostinelli:2002hh}
\hrefCMSnoop {} {{ GEANT4} Collaboration, ``{GEANT4}: A simulation toolkit'',}
  \textit{ Nucl. Instrum. Meth.} \textbf{ A506} (2003) 250.
\href{http://dx.doi.org/10.1016/S0168-9002(03)01368-8}{\texttt{
  doi:10.1016/S0168-9002(03)01368-8}}.

\bibitem{Maltoni:2002qb}
\hrefCMSnoop {} {F.~Maltoni and T.~Stelzer, ``{MadEvent}: Automatic event
  generation with {MadGraph}'',} \textit{ JHEP} \textbf{ 02} (2003) 027.
\href{http://dx.doi.org/10.1088/1126-6708/2003/02/027}{\texttt{
  doi:10.1088/1126-6708/2003/02/027}}.

\bibitem{Sjostrand:2007gs}
\hrefCMSnoop {} {T.~Sj{\"o}strand, S.~Mrenna, and P.~Skands, ``A Brief
  Introduction to {PYTHIA} 8.1'',} \textit{ Comput. Phys. Commun.} \textbf{
  178} (2008) 852.
\href{http://dx.doi.org/10.1016/j.cpc.2008.01.036}{\texttt{
  doi:10.1016/j.cpc.2008.01.036}}.

\bibitem{Nadolsky:2008zw}
\hrefCMSnoop {} {P.~M. Nadolsky {et~al.}, ``Implications of {CTEQ} global
  analysis for collider observables'',} \textit{ Phys. Rev.} \textbf{ D78}
  (2008) 013004, \href{http://www.arXiv.org/abs/0802.0007}{\texttt{
  arXiv:0802.0007}}.
\href{http://dx.doi.org/10.1103/PhysRevD.78.013004}{\texttt{
  doi:10.1103/PhysRevD.78.013004}}.

\bibitem{Khachatryan:2010ez}
\hrefCMSnoop {} {{ CMS} Collaboration, ``First Measurement of the Cross Section
  for Top-Quark Pair Production in Proton-Proton Collisions at $\sqrt{s}$=7
  {TeV}'',} \textit{ Phys. Lett.} \textbf{ B695} (2011) 424.
\href{http://dx.doi.org/10.1016/j.physletb.2010.11.058}{\texttt{
  doi:10.1016/j.physletb.2010.11.058}}.

\bibitem{Khachatryan:2010xn}
\hrefCMSnoop {} {{ CMS} Collaboration, ``Measurements of Inclusive {$W$} and
  {$Z$} Cross Sections in pp Collisions at $\sqrt{s}=7$ {TeV}'',} \textit{
  JHEP} \textbf{ 01} (2011) 1.
\href{http://dx.doi.org/10.1007/JHEP01(2011)080}{\texttt{
  doi:10.1007/JHEP01(2011)080}}.

\bibitem{Beenakker:1996ed}
\hrefCMSnoop {} {W.~Beenakker, R.~Hoepker, and M.~Spira, ``{PROSPINO}: A
  program for the Production of Supersymmetric Particles in Next-to-leading
  Order {QCD}'',} (1996).
  \href{http://www.arXiv.org/abs/hep-ph/9611232}{\texttt{
  arXiv:hep-ph/9611232}}.


\bibitem{Maruyama:1355424}
\href {http://cdsweb.cern.ch/record/1355424/files/CERN-THESIS-2011-023.pdf}
  {S.~Maruyama, ``Supersymmetry Search in Multilepton Final States with Taus
  Using the {CMS} Detector at the {LHC}''}.
\newblock PhD thesis, University of California at Davis, 2011.
\newblock {CERN-THESIS-2011-023}.

\bibitem{JME-10-010}
\href {http://cdsweb.cern.ch/record/1308178} {{ CMS} Collaboration,
  ``Determination of the Jet Energy Scale in {CMS} with pp Collisions at
  $\sqrt{s}= 7$ {TeV}'',} \textit{ CMS Physics Analysis Summary} \textbf{
  CMS-PAS-JME-10-010} (2010).

\bibitem{roostats}
\hrefCMSnoop {} {L.~Moneta {et~al.}, ``{The RooStats Project}'',} (2010).
  \href{http://www.arXiv.org/abs/1009.1003}{\texttt{ arXiv:1009.1003}}.

\bibitem{dmitry_hits_thesis}
\href {http://cdsweb.cern.ch/record/1363946/files/CERN-THESIS-2011-040.pdf}
  {D.~Hits, ``A multi-lepton search for new physics in $\mathrm{35~pb^{-1}}$
  proton-proton collisions at the {LHC} with a center mass energy of
  $\sqrt{s}=7$ {TeV} using the {CMS} detector''}.
\newblock PhD thesis, Rutgers, The State University of New Jersey, 2011.
\newblock {CERN-THESIS-2011-040}.

\bibitem{lepsusy}
\href
  {http://lepsusy.web.cern.ch/lepsusy/www/sleptons_summer04/slep_final.html} {{
  LEP2 SUSY Working Group} Collaboration, ``Combined {LEP} Selectron/Smuon/Stau
  Results, 183-208 {GeV}'',} (2004).

\bibitem{Heister:2001nk}
\hrefCMSnoop {} {{ ALEPH} Collaboration, ``Search for scalar leptons in $e^+
  e^-$ collisions at center-of-mass energies up to 209 {GeV}'',} \textit{ Phys.
  Lett.} \textbf{ B526} (2002) 206.
  \href{http://dx.doi.org/10.1016/S0370-2693(01)01494-0}{\texttt{
  doi:10.1016/S0370-2693(01)01494-0}}.

\bibitem{Heister:2003zk}
\hrefCMSnoop {} {{ ALEPH} Collaboration, ``Absolute mass lower limit for the
  lightest neutralino of the {MSSM} from $e^+ e^-$ data at $\sqrt{s}$ up to 209
  {GeV}'',} \textit{ Phys. Lett.} \textbf{ B583} (2004) 247.
  \href{http://dx.doi.org/10.1016/j.physletb.2003.12.066}{\texttt{
  doi:10.1016/j.physletb.2003.12.066}}.

\bibitem{Abdallah:2003xe}
\hrefCMSnoop {} {{ DELPHI} Collaboration, ``Searches for supersymmetric
  particles in $e^+ e^-$ collisions up to 208 {GeV} and interpretation of the
  results within the {MSSM}'',} \textit{ Eur. Phys. J.} \textbf{ C31} (2003)
  421. \href{http://dx.doi.org/10.1140/epjc/s2003-01355-5}{\texttt{
  doi:10.1140/epjc/s2003-01355-5}}.

\bibitem{Achard:2003ge}
\hrefCMSnoop {} {{ L3} Collaboration, ``Search for scalar leptons and scalar
  quarks at {LEP}'',} \textit{ Phys. Lett.} \textbf{ B580} (2004) 37.
  \href{http://dx.doi.org/10.1016/j.physletb.2003.10.010}{\texttt{
  doi:10.1016/j.physletb.2003.10.010}}.

\bibitem{Abbiendi:2003ji}
\hrefCMSnoop {} {{ OPAL} Collaboration, ``Search for anomalous production of
  dilepton events with missing transverse momentum in $e^+ e^-$ collisions at
  $\sqrt{s} =$ 183 - 209 {GeV}'',} \textit{ Eur. Phys. J.} \textbf{ C32} (2004)
  453. \href{http://dx.doi.org/10.1140/epjc/s2003-01466-y}{\texttt{
  doi:10.1140/epjc/s2003-01466-y}}.

\bibitem{D0sqgl}
\hrefCMSnoop {} {{ D0} Collaboration, ``Search for squarks and gluinos in
  events with jets and missing transverse energy using 2.1~$\mathrm{fb}^{-1}$
  of p$\mathrm{\bar{p}}$ collision data at {$\sqrt{s} = 1.96~\TeV$}'',}
  \textit{ Phys. Lett.} \textbf{ B660} (2008) 449.
  \href{http://dx.doi.org/10.1016/j.physletb.2008.01.042}{\texttt{
  doi:10.1016/j.physletb.2008.01.042}}.

\end{thebibliography}\endgroup
